\documentclass[12pt]{article}

\AtBeginDocument{
  \addtocontents{toc}{\normalsize}
}

\pdfoutput=1

\usepackage{amsmath,amssymb,amsfonts,amsthm,amscd,mathrsfs}
\usepackage{xcolor}
\definecolor{darkblue}{rgb}{0.1,0.1,.7}
\usepackage[]{version}
\usepackage[]{graphicx}
\usepackage[]{latexsym}
\usepackage{geometry}
\usepackage[all,cmtip]{xy}
\usepackage[margin=10pt,font=small,labelfont=bf]{caption}
\usepackage{ifthen}
\usepackage{tikz}
\usepackage{array,setspace,mathrsfs,amsfonts,yfonts,dsfont,bbm,colonequals}
\usepackage{dsfont}
\usepackage{cite}
\usepackage{xspace}

\usepackage{lscape}
\usepackage{pdflscape}
\usepackage{appendix}
\usepackage[english]{babel}
\usepackage[utf8x]{inputenc}
\usepackage[T1]{fontenc}
\usepackage{slashed}
\usepackage{braket}
\usepackage{amsmath,amsthm}
 
\theoremstyle{definition}

\theoremstyle{plain}

\usepackage{amsmath}
\usepackage{mathtools}
\usepackage{amsthm}
\usepackage{amssymb}
\usepackage{amsfonts}
\usepackage{graphicx}
\usepackage[colorinlistoftodos]{todonotes}
\usepackage[colorlinks=true, allcolors=blue]{hyperref}
\numberwithin{equation}{section}

\newcommand{\mt}[1]{(\mbox{\tiny #1})}

\newcommand{\reef}[1]{(\ref{#1})}
\newcommand{\be}{\begin{equation}}
\newcommand{\ee}{\end{equation}}
\newcommand{\bea}{\begin{eqnarray}}
\newcommand{\eea}{\end{eqnarray}}
\newcommand{\ba}{\begin{equation}\begin{aligned}}
\newcommand{\ea}{\end{aligned}\end{equation}}

\begin{document}

\vspace*{-.6in} \thispagestyle{empty}
\begin{flushright}
LPTENS/18/08
\end{flushright}
\vspace{1cm} {\Large
\begin{center}
{\bf Bounding scattering of charged particles in 1+1 dimensions}\\
\end{center}}
\vspace{1cm}
\begin{center}
{\bf Miguel F.~Paulos, Zechuan Zheng}\\[1cm] 
{
\small
{\em Laboratoire de Physique Th\'eorique de l'\'Ecole Normale Sup\'erieure\\ PSL University, CNRS, Sorbonne Universit\'es, UPMC Univ. Paris 06\\ 24 rue Lhomond, 75231 Paris Cedex 05, France
}
\normalsize
}
\\
\end{center}

\vspace{4mm}

\begin{abstract}
We obtain general bounds on scattering processes involving charged particles in 1+1 spacetime dimensions. After a general analysis we derive mostly numerical bounds on couplings in theories with $O(N)$ and $U(N)$ global symmetries. The bounds are consistently saturated by $S$-matrices without particle production, and in many cases by known integrable $S$-matrices. Our work provides a blueprint for a similar analysis in higher dimensions.

\end{abstract}
\vspace{2in}


\newpage

{
\setlength{\parskip}{0.05in}
\tableofcontents
\renewcommand{\baselinestretch}{1.0}\normalsize
}


\setlength{\parskip}{0.1in}
\newpage

\section{Introduction}
The study of scattering processes in 1+1 dimensions has led to a wealth of exact results in the context of integrable models \cite{mussardo2010statistical,Dorey1996b,1979AnPhy.120..253Z}. Exact $S$-matrices can be found using unitarity, crossing symmetry, analyticity and the Yang-Baxter equation, under the guise of factorized scattering. However, it has been long realized (though not widely appreciated \cite{Creutz1972a}) that the first three of these assumptions already tell us much about the properties of general quantum field theories (QFTs). 

Previous work \cite{Paulos2017} obtained general bounds on scattering processes involving scalar particles. Consider a gapped quantum field theory in 1+1 dimensions, and consider the 2-to-2 $S$-matrix describing scattering of the lightest scalar particle. Assuming these particles can exchange a fixed number of one or more bound states, we can ask: can the coupling to a particular bound state be arbitrarily large\footnote{See \cite{Doroud2018a} for a generalization to resonances.}? Physically one expects a definite ``no'', since increasing the coupling should eventually give rise to the appearance of new bound states. To answer this question, one first comes up with an ansatz that takes full advantage of the analyticity and crossing properties of $S$-matrices. One then maximizes the coupling (numerically or analytically) subject to the unitarity constraints.

The goal of this note is to explain how it is possible to obtain more constraining bounds under the assumption that the particles involved in a given scattering process transform in an irreducible representation of a global symmetry group $G$. Although our kinematical analysis will be general, for the purpose of obtaining concrete (numerical) bounds we will focus on two particular cases, namely those of particles transforming in the vector and fundamental representations of the $O(N)$ and $U(N)$ groups respectively. We find that our bounds are systematically saturated by $S$-matrices without any particle production. Known integrable models, such as the Gross-Neveu models with $O(N)$ and $U(N)$ symmetry, as well as the sine-Gordon model, saturate the bounds. We find it remarkable that highly non-trivial $S$-matrices with complicated analytic structure can be reproduced numerically in this way with high precision.
However, unlike the uncharged case, we have not been able to find a closed form solution for the optimal $S$-matrices. It would be very interesting to attempt to derive them analytically from an optimization principle.

The layout of this note is as follows. In section \ref{sec:kin} we discuss the general setup, including the kinematics and group theory analysis. We then focus on the two particular cases that will be relevant, namely scattering of $O(N)$ and $U(N)$ vector particles. We will briefly review the associated integrable models in section \ref{sec:review}. In section \ref{sec:pole} we describe the optimization problem we interested in and make a few analytic observations. The numerical analysis and results are then presented in section \ref{sec:num}. Although our bounds hold for generic gapped QFTs, we observe that in many cases they are saturated by known integrable $S$-matrices. This note is complemented by appendices containing further numerical results. 
\vspace{0.3cm}

{\em Note:} While this note was being completed the work \cite{yifei2018} appeared which overlaps with ours, and we also became aware of similar work to appear by C\'ordova and Vieira.

\section{Kinematics of charged $S$-matrices}
\label{sec:kin}
In this section we will overview the kinematics 
of scattering processes involving particles charged under a global symmetry. We concentrate on the case of $2\rightarrow 2$ scattering in $1+1$ dimensions. The $S$-matrix is defined as:
\bea
\mathcal S&=&\sum_{\rho} |\rho, \mathrm{in}\rangle\,\, \!\langle \rho, \mathrm{out}|,
\eea
where the states $|\rho, \mathrm{in}\rangle$ and $\langle \rho, \mathrm{out}|$ are respectively asymptotic incoming and outgoing particle states, with the schematic label $\rho$ standing for the associated quantum numbers. Both in and out states form a complete basis of the physical Hilbert space, with the $S$-matrix a unitary operator which maps us from one basis to the other. The normalization of one-particle asymptotic states is taken to be:
\begin{equation}
\Braket{p_j,\sigma_j,\mathrm{in}|p_i,\sigma_i,\mathrm{in}}=\Braket{p_j,\sigma_j,\mathrm{out}|p_i,\sigma_i,\mathrm{out}}=\bar \delta_{ij}\delta_{\sigma_{i},\sigma_j} \equiv (2\pi)\,2E_i\,\delta(k_i-k_j)\delta_{\sigma_{i},\sigma_j}
\end{equation}
where $\sigma$ stands for the remaining quantum numbers of the particle and the momenta are $p^\mu=(E,k)$. From now on we will drop the in and out labels since they should be clear from context.

We will be focusing on 2-to-2 scattering of particles with the same mass $m$ (which we will eventually set to 1). The particular $S$-matrix elements we will be interested in take the form
\begin{equation}\label{smatrix}
\Braket{ p_4,\sigma_4; p_3,\sigma_3|\mathcal S| p_1,\sigma_1;p_2, \sigma_2}=
F_{\sigma_4 \sigma_3,\sigma_1 \sigma_2}(s) \bar \delta_{13}\bar\delta_{24}+ R_{\sigma_3 \sigma_4,\sigma_2 \sigma_1}(s)\bar \delta_{14}\bar \delta_{23}
\end{equation}
We have written the $S$-matrix element in terms of forward $F$and reflected $R$ amplitudes. Lorentz invariance tells us that the amplitude can only depend on the Mandelstam invariants,
\bea
s=-(p_1+p_2)^2,\qquad t=-(p_1-p_4)^2,\qquad u=-(p_1-p_3)^2
\eea
subject to the Mandelstam relation,
\bea
s+t+u=4m^2.
\eea
In $1+1$ dimensions we must have $t u=0$ which means the $S$-matrix elements can be expressed in terms of the single invariant $s$. In practice it is useful to work on the rapidity plane by introducing the $\theta$ variable:
\bea
s=4m^2\, \cosh(\theta/2)^2.
\eea
Physical scattering processes take place for $s$ larger than $4m^2$, with a slightly positive imaginary part imposed by the Feynman $i\epsilon$ prescription. However, the $S$-matrix can be analytically continued off the physical region and onto the complex plane. In particular scattering processes in different channels may be obtained by analytic continuation - this is called crossing symmetry. Since we are scattering the lightest particle in the theory, the possible singularities consist of poles for $0<s<4m^2$, which describe physical bound states, the physical region cut for $s>4m^2$ and similar singularities obtained from crossing symmetry from other $S$-matrix elements.  

After these general remarks, we will now consider the case where the particles being scattered transform as irreducible representations of a global symmetry group $G$.
We will work out the result for the cases where $G$ is real or complex separately, keeping in mind our desired applications to $O(N)$ and $U(N)$. 

\subsection{The real group case and application to $O(N)$}

Consider the case where all particles transform in some irreducible representation $\mathcal R$ of the real group $G$. Since we are interested in scattering states containing two charged particles, we are led to consider the tensor product decomposition:
\bea
\mathcal R\otimes \mathcal R=\bigoplus_{i} \mathcal P_i
\eea
In terms of states we have
\bea
|\mathcal R,\alpha\rangle\otimes |\mathcal R,\beta\rangle=\sum_{i}\sum_{n=1}^{M_i} \sum_{\rho=1}^{d_i} \left( C^{(i)}_{n}\right)_{\alpha\beta,\rho}\, |\mathcal P_i,\rho\rangle
\eea
where $d_i$ is the dimension of the representation $\mathcal P_i$, $M_i$ the multiplicity with which it appears in the tensor product $\mathcal R\otimes \mathcal R$
and $C_n^{(i)}$ the associated Clebsch-Gordan coefficients which can be chosen real. Here $\alpha,\beta,\rho$ label individual basis elements in the vector space of the corresponding representations. For what concerns the group structure, the $S$-matrix can be written as
\bea
\langle \delta ,\gamma| \mathcal S| \alpha ,\beta\rangle=\sum_{i,j} \sum_{n,m=1}^{M_i} \sum_{\lambda=1}^{d_j}\sum_{\rho=1}^{d_i} \left( C^{(i)}_m\right)_{\gamma \delta,\rho} \left( C^{(j)}_n\right)_{\alpha\beta,\lambda} \langle \mathcal P_j,m,\rho| \mathcal S| \mathcal P_i,n,\lambda\rangle\,.
\eea
Schur's lemma now implies the important result:
\bea
\langle \mathcal P_j,m,\rho| \mathcal S| \mathcal P_i,n,\lambda\rangle= \delta_{ij} \delta_{\rho\lambda}\, S^{(i)}_{mn}.
\eea
Since we are scattering indistinguishable particles transforming in a real representation, we may set in \reef{smatrix}:
\bea
F_{\delta \gamma,\alpha\beta}=\pm R_{\gamma\delta,\alpha\beta}\equiv S_{\delta \gamma,\alpha\beta}
\eea
where $+$ and $-$ signs correspond to scattering of bosons or fermions respectively. Using the expression for the S-matrix written above we can write
\bea
S_{\delta \gamma,\alpha\beta}(\theta)=\sum_{i}\sum_{m,n=1}^{M_i} \left(T^{(i)}_{mn}\right)_{\delta \gamma,\alpha \beta}\, S^{(i)}_{mn} (\theta)\,,
\eea
with the $G$ invariant tensors
\bea
  \left(T^{(i)}_{mn}\right)_{\delta \gamma,\alpha \beta}=\sum_{\lambda=1}^{d_i} \left( C^{(i)}_m\right)_{\gamma\delta,\lambda} \left( C^{(i)}_{n}\right)_{\alpha\beta,\lambda}.
\eea
Physically we have decomposed the total 2-to-2 $S$-matrix as a sum of ``partial waves'', or channels, with definite transformation properties under the group $G$.

The unitarity condition on the $S$-matrix states that
\bea
%
\langle k_4,\delta; k_3,\gamma| \mathcal S^\dagger \mathcal S |k_1,\alpha; k_2,\beta\rangle =\bar \delta_{13}\bar \delta_{24} \delta_{\alpha\gamma}\delta_{\beta \delta}\pm\bar \delta_{14}\bar \delta_{23} \delta_{\alpha\delta}\delta_{\beta \gamma}.
\eea
Using orthogonality of the Clebsch-Gordan coefficients this implies:
\bea
\mathds 1-\left[\left(S^{(i)}\right)^{\dagger}\cdot S^{(i)}\right](\theta)
\succeq 0, \qquad \mbox{for}\quad \mathcal P_i \subset \mathcal R\otimes \mathcal R, \qquad \theta \in \mathbb R.
\eea
where $\mathds 1$ is the identity matrix of rank $M_i$. In other words, unitarity becomes diagonal in the partial wave decomposition, with each $S$-matrix separately satisfying positive semidefiniteness conditions. In the case where the multiplicity is one this reduces to the familiar constraint
\bea
|S^{(i)}(s)|^2\leq 1, \qquad s\geq 4m^2.
\eea
In this paper we will be interested in the case where $\mathcal R$ is the $N$ dimensional vector representation of $O(N)$. In this case we have the tensor product decomposition
\bea
N\otimes N=1\oplus \frac{N(N+1)-2}2 \oplus \frac{N(N-1)}2.
\eea
which correspond to the singlet, symmetric traceless tensor, and antisymmetric tensor representations respectively. We will  denote these by $(s),(t)$ and $(a)$. The basis of invariant tensors is given by
\begin{subequations}
\bea
T^{(s)}_{\delta \gamma,\alpha \beta}&=&\frac{1}{N} \delta_{\alpha \beta}\delta_{\delta\gamma}\\
T^{(t)}_{\delta \gamma,\alpha \beta}&=&\frac{1}{2}(\delta_{\alpha\delta}\delta_{\beta\gamma}+\delta_{\alpha\gamma}\delta_{\beta\delta})-\frac{1}{N} \delta_{\alpha \beta}\delta_{\delta \gamma}\\
T^{(a)}_{\delta \gamma,\alpha \beta}&=&\frac{1}{2}(\delta_{\alpha\gamma}\delta_{\beta\delta}-\delta_{\alpha\delta} \delta_{\beta\gamma})
\eea
\end{subequations}
Using this basis we can decompose the $S$-matrix into the three physical channels corresponding to propagation of the $(s),(t)$ and $(a)$ representations:
\bea
S_{\delta\gamma ,\alpha \beta}(\theta)=\sum_{i \in \{s,t,a\}}\, T^{(i)}_{\delta \gamma,\alpha \beta}\, S^{(i)}(\theta)
\eea
Completeness of the basis of invariant tensors implies the crossing relations:
\bea
T^{(i)}_{\beta \gamma ,\alpha \delta}=\sum_{j\in\{s,t,a\}} \mathcal F^{(i,j)}\, T^{(j)}_{\delta \gamma,\alpha \beta}, \qquad S^{(i)}(\theta)=\sum_{j\in\{s,t,a\}} \mathcal F^{(j,i)}S^{(j)}(i\pi-\theta)
\eea
where the tensor $\mathcal F$ can be explicitly computed:
\bea
\mathcal F^{(i,j)}
=\frac{1}{2N}\left(\begin{array}{ccc}
2 & 2&-2\\
N(N+1)-2&N-2 &N+2\\
-N(N-1)&N&N.
\end{array}
\right)_{ij}, \qquad i,j \in \{s,t,a\}.
\eea
Recall that crossing symmetry reflects the fact that the same analytic $S$-matrix can describe scattering in different channels. In the above, crossing describes how to go from scattering of particles $1,2\to 3,4$ to a process $1,4\to 3,2$.
It turns out to be convenient to introduce a different basis, writing
\bea
S_{\delta \gamma ,\alpha \beta}^{}(\theta)=\delta_{\alpha \beta}\delta_{\gamma \delta} S_1(\theta)+\delta_{\alpha\delta} \delta_{\beta\gamma} S_2(\theta)+\delta_{\alpha\gamma} \delta_{\beta\delta} S_3(\theta).
\eea
We have the identifications
\bea
S^{(s)}&=& N S_1+ S_2+S_3, \\
S^{(t)}&=& S_2+S_3,\\
S^{(a)}&=& S_2-S_3.
\eea
In this basis the statement of crossing symmetry becomes very simple, namely
\bea
S_2(\theta)=S_2(i\pi-\theta),\qquad S_1(i\pi -\theta)=S_3(\theta).
\eea

\subsection{The complex group case and application to $U(N)$}
We now consider the case where we have a complex symmetry group. We are interested in a scattering process involving particles transforming under a complex representation $\mathcal R$ and antiparticles transforming in $\overline{\mathcal R}$. Note that we define
\bea
\left(|\mathcal R,\alpha\rangle\right)^\dagger=\langle \overline{\mathcal R}, \bar \alpha|.
\eea
We now have the tensor product decompositions
\bea
\mathcal R\otimes \overline{\mathcal R}=\bigoplus_i \mathcal Q_i(+\overline{\mathcal Q_i}), \qquad \mathcal R\otimes \mathcal R=\bigoplus_i \mathcal P_i,
\eea
where in the first decomposition we simply noted that if a particular complex representation appears, so must its complex conjugate.

The 2-to-2 $S$-matrices for  particle-particle and particle-antiparticle scattering respectively can be expressed as
\bea\label{smatrixcomplex}
\Braket{ p_4,\bar \delta; p_3,\bar \gamma|\mathcal S| p_1,\alpha;p_2, \beta}&=&
S_{\bar \delta\bar \gamma,\alpha \beta}(\theta)\left[\bar \delta_{13}\bar \delta_{24}\pm \bar \delta_{14}\bar\delta_{23}\right]\\
\Braket{ p_4,  \delta; p_3,\bar \gamma|\mathcal S| p_1,\alpha;p_2, \bar \beta}&=&
F_{\delta \bar \gamma,\alpha \bar \beta}(\theta) \bar \delta_{13}\bar\delta_{24}+ R_{\bar \gamma \delta ,\alpha \bar \beta}(\theta)\bar \delta_{14}\bar \delta_{23}
\eea
where the positive (negative) sign is suitable for bosonic (fermionic) scattering.
Crossing symmetry now requires: 
\bea
S_{\bar \delta \bar \gamma,\alpha\beta}(i\pi-\theta)=F_{\beta \bar \gamma,\alpha \bar \delta}(\theta)\,,\\
R_{\bar\gamma \delta,\alpha \bar \beta}(i\pi-\theta)=R_{\bar\gamma \alpha,\delta \bar \beta}(\theta).
\eea
In what follows it is convenient to work with combinations of $F,R$ with definite transformation properties under parity transformations. Accordingly we define
\bea
F^{\pm}_{\delta \bar \gamma,\alpha \bar \beta}=F_{\delta \bar \gamma,\alpha \bar \beta}\pm R_{\bar\gamma \delta,\alpha \bar \beta}.
\eea
where the sign now denotes parity and is uncorrelated with the previous one.
The discussion now proceeds as before, except that we have two sets of invariant tensors. For simplicity, consider the case where $\mathcal P_i,\mathcal Q_j$ appear with unit multiplicity in $\mathcal R\otimes \bar{ \mathcal  R}$ and $\mathcal R\otimes \mathcal R$ respectively. Then we have
\bea
S_{\bar \delta \bar \gamma,\alpha\beta}=\sum_{\mathcal P_i} T^{(i)}_{\bar \delta \bar \gamma,\alpha\beta}\, S^{(i)},\qquad T^{(i)}_{\bar \delta \bar \gamma,\alpha\beta}=\sum_{\lambda=\bar\lambda=1}^{d_i}  C^{(\bar i)}_{\bar \gamma\bar \delta,\bar \lambda}  C^{(i)}_{\alpha\beta,\lambda},\\
F^{\pm}_{\delta \bar \gamma,\alpha\bar \beta}=\sum_{\mathcal Q_i} T^{(i)}_{\delta  \bar \gamma,\alpha\bar \beta}\, S^{(i)_{\pm}},\qquad T^{(i)}_{\delta  \bar \gamma,\alpha\bar \beta}=\sum_{\lambda=\bar\lambda=1}^{d_i}  C^{(\bar i)}_{\bar \gamma \delta,\bar \lambda}  C^{(i)}_{\alpha\bar \beta,\lambda}.
\eea
Completeness of the invariant tensor basis implies once again that there exist crossing matrices $\mathcal F,\tilde{\mathcal F}$ such that
\bea
T^{(i)}_{\delta  \bar \beta,\alpha\bar \gamma}=\sum_{\mathcal Q_j} \mathcal F^{(i,j)}T^{(j)}_{\delta  \bar \gamma,\alpha\bar \beta},\qquad T^{(i)}_{\bar \gamma \beta ,\alpha \bar \delta}=\sum_{\mathcal P_j} \tilde{\mathcal F}^{(i,j)} T^{(j)}_{\bar \delta  \bar \gamma,\alpha \beta}.
\eea

We now focus on the case where $\mathcal R$ is the fundamental representation of $U(N)$. In this case we have
\bea
N\otimes \bar N&=&1\oplus (N^2-1),\\
N\otimes N&=& \frac{N(N-1)}2\oplus \frac{N(N+1)}2
\eea
The representations appearing on the first line are the singlet and adjoint representations, and on the second line the symmetric and antisymmetric tensor representations. The full set of invariant tensors is given by
\begin{subequations}
\begin{align}
T^{(\mbox{\tiny sing})}_{\delta \bar \gamma,\alpha\bar\beta}&=\frac 1N\,\delta_{\alpha \bar \beta} \delta_{\delta \bar \gamma},& T^{(\mbox{\tiny adj})}_{\delta \bar \gamma,\alpha\bar\beta}&=\delta_{\alpha \bar \gamma} \delta_{\delta \bar \beta}-\frac 1N\,\delta_{\alpha \bar \beta}\delta_{\delta \bar \gamma}\\
T^{(\mbox{\tiny sym})}_{\bar \delta \bar \gamma,\alpha\beta}&=\frac 12\left(\delta_{\alpha \bar \gamma} \delta_{\beta \bar \delta}+\delta_{\alpha \bar \delta} \delta_{\beta \bar \gamma}\right),& T^{(\mbox{\tiny asym})}_{\bar \delta \bar \gamma,\alpha\beta}&=\frac 12\left(\delta_{\alpha \bar \gamma} \delta_{\beta \bar \delta}-\delta_{\alpha \bar \delta} \delta_{\beta \bar \gamma}\right).
\end{align}
\end{subequations}
Accordingly the crossing matrices become
\begin{align}
\mathcal F^{(i,j)}&=\frac{1}{N}\left(\begin{array}{cc}
1&1\\N^2-1& -1\end{array}\right),& i,j &\in \{\mbox{sing},\mbox{adj}\}\\
\tilde{\mathcal F}^{(i,j)}&=\frac{1}{N}\left(\begin{array}{cc}
1&-1\\N-1& N+1\end{array}\right) ,& i &\in \{\mbox{sing},\mbox{adj}\},\quad j\in\{\mbox{sym},\mbox{asym}\}.
\end{align}
That is, we have
\begin{subequations}
\bea
S^{(i)_+}(\theta)-S^{(i)_-}(\theta)&=&\!\!\!\sum_{j\in\{\mbox{\tiny sing},\mbox{\tiny adj}\}}\!\! \mathcal F^{(j,i)} \left[S^{(j)_+}(i\pi-\theta)-S^{(j)_-}(i\pi-\theta)\right], \\
S^{(i)_+}(\theta)+S^{(i)_-}(\theta)&=&\!\!\!\sum_{j\in\{\mbox{\tiny sym},\mbox{\tiny asym}\}}\!\! \left(\tilde{\mathcal F}^{-1}\right)^{(j,i)} S^{(j)}(i\pi-\theta),\\
S^{(i)}(\theta)&=&\!\!\!\sum_{j\in\{\mbox{\tiny sing},\mbox{\tiny adj}\}}\!\! \tilde{\mathcal F}^{(j,i)}\left[ S^{(j)_+}(i\pi-\theta)+S^{(j)_-}(i\pi-\theta)\right], 
\eea
\end{subequations}
where the first two equations hold for $i\in \{\mbox{sing},\mbox{adj}\}$ and the second for $i \in\{\mbox{sym},\mbox{asym}\}$.
The unitarity conditions in each channel are:
\bea
|S^{(\mbox{\tiny sing})_\pm}(s)|^2\leq 1,\qquad |S^{(\mbox{\tiny adj})_\pm}(s)|^2\leq 1,\\
|S^{(\mbox{\tiny sym})}(s)|^2\leq 1,\qquad |S^{(\mbox{\tiny asym})}(s)|^2\leq 1,
\eea
for $s>4m^2$.
The crossing properties are particularly simple in a different tensor basis. Following Berg et al \cite{1978NuPhB.134..125B} we define:
\begin{subequations}
\label{eq:undecomp}
\begin{align}
S^{\mt{sym}}&=u_1+u_2,& S^{\mt{asym}}&=u_1-u_2\\
S^{\mt{sing}_{\pm}}&=t_1\pm r_1+N(t_2\pm r_2),&\qquad S^{\mt{adj}_\pm}&=t_1\pm r_1.
\end{align}
\end{subequations}
Then crossing symmetry becomes
simply:
\bea\label{uncross}
u_1(\theta)=t_1(i\pi-\theta)\qquad 
u_2(\theta)=t_2(i\pi-\theta) \qquad 
r_1(\theta)=r_2(i\pi-\theta).
\eea

\section{Review of integrable $S$-matrices}
\label{sec:review}
We will be interested in deriving upper bounds on couplings appearing in 2-to-2 $S$-matrices. Past experience \cite{Creutz1972a,Paulos2017} leads us to expect that these bounds are generically saturated by $S$-matrices without any particle production.
A simple explanation for this is that bounds exist only because there are constraints, namely unitarity, and so at the bound as many constraints will be saturated as possible. Hence the associated $S$-matrices should also saturate unitarity. Before we embark on our numerical explorations it is worthwhile to review what kinds of such $S$-matrices are known to exist, namely in integrable models. We split our short review into the two cases of interest, namely $O(N)$ and $U(N)$.

\subsection{$O(N)$ $S$-matrices}
In \cite{1979AnPhy.120..253Z} the authors constructed
$S$-matrices describing 2-to-2 scattering of particles charged under $O(N)$, that satisfy crossing symmetry, unitarity and the Yang-Baxter equation. 
The minimal $O(2)$ solution corresponds to the $S$-matrix of the sine-Gordon model which describes scattering of solitons. For $N\geq 3$ there are two classes of minimal $S$-matrices which are believed to correspond to the $O(N)$ non-linear sigma model, and the Gross-Neveu model.
\subsubsection*{$O(2)$: sine-Gordon model}
The Lagrangian of the sine-Gordon Model is:
\begin{equation}
\mathcal{L}=\frac{1}{2} (\partial_{\mu} \phi)^2+\frac{m^2}{\beta^2}(\cos \beta \phi -1)
\end{equation}
In \cite{coleman1975quantum,mandelstam1975soliton}, it was first argued that the sine-Gordon model is equivalent to the massive Thirring model:
\begin{equation}
\mathcal{L}_{MTM}=i\bar{\psi}\gamma^\mu \partial_\mu \psi-m_0\bar{\psi}\psi-\frac{g}{2}(\bar{\psi \gamma^\mu \psi})^2,
\end{equation}
with the soliton of the sine-Gordon model identified as the elementary fermion in the Thirring model. In the latter description it is clear that there is a $O(2)\cong U(1)$ symmetry (fermion number). It is convenient to introduce:
\begin{equation}
\xi=\frac{\beta^2}{8}\frac{1}{1-\frac{\beta^2}{8\pi}}
\end{equation}
as the renormalized coupling constant. 
The $S$-matrices for scattering of solitons in the $O(2)$ language are \cite{mussardo2010statistical}:
\begin{equation}
\begin{split}
	S^{(t)}(\theta)=&-\prod^{\infty}_{k=0}\frac{\Gamma\left(1+(2k+1)\frac{\pi}{\xi}-i\frac{\theta}{\xi}\right)\Gamma\left(1+2k\frac{\pi}{\xi}+i\frac{\theta}{\xi}\right)}{\Gamma\left(1+(2k+1)\frac{\pi}{\xi}+i\frac{\theta}{\xi}\right)\Gamma\left(1+2k\frac{\pi}{\xi}-i\frac{\theta}{\xi}\right)}\\
	&\times \frac{\Gamma\left((2k+1)\frac{\pi}{\xi}-i\frac{\theta}{\xi}\right)\Gamma\left((2k+2)\frac{\pi}{\xi}+i\frac{\theta}{\xi}\right)}{\Gamma\left((2k+1)\frac{\pi}{\xi}+i\frac{\theta}{\xi}\right)\Gamma\left((2k+2)\frac{\pi}{\xi}-i\frac{\theta}{\xi}\right)}
\end{split}
\end{equation}
\begin{equation}
S^{(s)}(\theta)+S^{(a)}(\theta)=\frac{2\sinh \frac{\pi \theta}{\xi}}{\sinh \frac{\pi (i\pi-\theta)}{\xi}}S^{(t)}(\theta)
\end{equation}
\begin{equation}
S^{(s)}(\theta)-S^{(a)}(\theta)=i\frac{2\sin \frac{\pi^2}{\xi}}{\sinh \frac{\pi (i\pi-\theta)}{\xi}}S^{(t)}(\theta)
\end{equation}
When scattering solitons we can generate bound states called breathers which appear as poles in the $S$-matrix at specific values of $\theta$, namely
\begin{equation}
\theta_n=i \pi-i n\xi, \qquad \textrm{for}\quad n=1,\ldots \left\lfloor \frac{\pi}{\xi} \right\rfloor. 
\end{equation}
as well as cross channel poles at $\theta=i n\xi$. It is easy to check that breathers with $n$ even/odd correspond respectively to scalar/antisymmetric tensor particles. Since $N=2$ the latter can also be thought of as pseudoscalar particles. As for the $(t)$ channel $S$-matrix, it only contains poles related to the previous ones by crossing symmetry.

\subsubsection*{$O(N)$ with \texorpdfstring{$N\geq 3$}-}
In this case there are now two minimal solutions for the $S$-matrix:

\begin{equation}
S_2^\pm(\theta) = Q^{\pm}(\theta)Q^{\pm}(i\pi-\theta)
\end{equation}
with:
\begin{equation}
S_3=-\frac{i\lambda}{\theta} S_2(\theta),\qquad
S_1=-\frac{i\lambda}{i\pi-\theta} S_2(\theta)
\end{equation}
where $\lambda=\frac{2\pi}{N-2}$ and:
\begin{equation}
Q^\pm(\theta)=\frac{\Gamma(\pm\frac{\lambda}{2\pi}-i\frac{\theta}{2\pi})\Gamma(\frac{1}{2}-i\frac{\theta}{2\pi})}{\Gamma(\frac{1}{2}\pm\frac{\lambda}{2\pi}-i\frac{\theta}{2\pi})\Gamma(-i\frac{\theta}{2\pi})}.
\end{equation}
There is strong evidence that the plus sign corresponds to $O(N)$ symmetric non-linear sigma model and minus sign corresponds to $O(N)$ symmetric Gross-Neveu model, with Lagrangians given respectively by
\bea
\mathcal{L}_{\sigma}&=&\frac{1}{2g}\sum^N_{i=1}(\partial_\mu n_i)^2 \qquad \mbox{with}\qquad\sum_{i=1}^N n_i^2=1,\\
\mathcal{L}_{GN}&=&\frac{i}{2} \sum^N_{i=1} \bar{\psi}_i \gamma_\mu \partial_\mu \psi_i +\frac{g}{8} \left[\sum^N_{i=1} \bar{\psi}_i \psi_i\right]^2.
\eea
In the latter $\psi_i$ are Majorana fermions.

There is no physical bound state for the first minimal solution, and in particular for the non-linear sigma model. In the cases $N=3,4$ the two $S$-matrices turn out to be the same, and in particular there is no bound state for these cases. However, when $N\geq 5$, we have the relation
\bea
S_2^{-}(\theta)=S_2^{+}(\theta)\times \frac{\sinh\theta+i\sin\lambda}{\sinh\theta-i\sin\lambda}
\eea
In particular the $S$-matrix describing scattering of the elementary fermion of the Gross-Neveu model contains a bound state at $\theta=i \lambda$. This corresponds physically to $s$-channel poles in the scalar and antisymmetric tensor channels with identical masses. The model contains other states, but these do not appear in the particular $S$-matrix elements that we are considering. However, we should point out some peculiar features. For $N=6$ the pole actually becomes a double pole and for $N=5$ it has the incorrect sign for the residue. Hence to get a physical $S$-matrix it seems we can multiply it by an overall minus sign, but whether this leads to an overall consistent theory is not clear.

\subsection{$U(N)$ $S$-matrices}

Following the work of Zamolodchikov and Zamolodchikov \cite{1979AnPhy.120..253Z}, B. Berg et al. \cite{1978NuPhB.134..125B} classified the minimal solutions for $S$-matrices with $U(N)$ symmetry. These minimal solutions fall into six classes listed in tables \ref{IIV} and \ref{VVI}. Here we are using the notation introduced in equations \reef{eq:undecomp}, and those functions which are unlisted may be obtained using the crossing relations \reef{uncross}. We have introduced the variable $\hat \theta=\frac{\theta}{i\pi}$and $f(\hat \theta,\lambda)$ is defined as:
\begin{equation}
f(\hat \theta,\lambda)=\frac{\Gamma(\frac{1}{2}+\frac{1}{2}\hat \theta)\Gamma(\frac{1}{2}+\frac{1}{2}\lambda-\frac{1}{2}\hat \theta)}{\Gamma(\frac{1}{2}-\frac{1}{2}\hat \theta)\Gamma(\frac{1}{2}+\frac{1}{2}\lambda+\frac{1}{2}\hat \theta)}
\end{equation}

\begin{table}
	\centering
	\begin{tabular}{cccccccccccccc}
	\hline
		Class 	&Parameter	&$t_1(\hat \theta)$	 	&$r_1(\hat \theta)$		&$t_2(\hat \theta)$\\
	\hline
	I	&	&1	&0	&0\\
	II	&$\lambda=\frac{2}{N}$	&$f(\hat \theta,\lambda)$		&0		&$-\frac{\lambda}{1-\hat \theta}t_1(\hat \theta)$\\
	III	&$\lambda=\frac{1}{N-1}$&$f(\hat \theta,\lambda)f(1-\hat \theta,\lambda)$&$-\frac{\lambda}{\hat \theta}t_1(\hat \theta)$	&$-\frac{\lambda}{1-\hat \theta}t_1(\hat \theta)$	\\
	IV	&$\lambda=\frac{1}{N+1}$&$-f(\hat \theta,\lambda)f(1-\hat \theta,\lambda)\tan (\frac{\pi}{2}(\hat \theta+\frac{1}{2}))$&$\frac{\lambda}{\hat \theta}t_1(\hat \theta)$&$-\frac{\lambda}{1-\hat \theta}t_1(\hat \theta)$	\\
	\hline
	\end{tabular}
	\caption{\label{IIV}Classes I-IV of $U(N)$ minimal solutions}
\end{table}

\begin{table}
	\centering
	\begin{tabular}{cccccccccccccc}
	\hline
		Class 	&Parameter	&$t_1(\hat \theta)$	 	&$r_1(\hat \theta)$		&$t_2(\hat \theta)$\\
	\hline
	V	&$\cosh(\pi\mu)=N$		&0		&${\displaystyle \prod_{k=-\infty}^{\infty} \frac{f(\hat \theta,k/(2i\mu))}{f(\hat \theta,k/(2i\mu)+1/2)}}$&$r_1(1-\hat \theta)$\\
	VI	&$e^{\pi \mu}=N$			&0		&${\displaystyle \prod_{k=-\infty}^{\infty} \frac{f(\hat \theta,k/(2i\mu))}{f(\hat \theta,k/(2i\mu)+1/2)}}$&$N^{-(1-\hat \theta)}r_1(1-\hat \theta)$\\
	\hline
	\end{tabular}
	\caption{\label{VVI}Classes V and Class VI of $U(N)$ minimal solutions. Note that the parameter $\mu$ is a real number.}
\end{table}
Note also in the tables we have abused notation by writing e.g. $r_1(\theta(\hat \theta))=r_1(\hat \theta)$.
We now examine these solutions in turn. Class I is trivial. In a series of papers\cite{berg1978exact,koberle1979scattering,abdalla1979more}, the Class II solution has been identified with the $SU(N)$ Gross-Neveu model\footnote{The model defined by Eq. \ref{UGNL} is also called chiral $SU(N)$.}\cite{gross1974dynamical}:

\begin{equation}\label{UGNL}
\mathcal{L}=i\sum_{i=1}^N \bar{\psi}_i \slashed{\partial} \psi_i +\frac{1}{2} g^2\left[\left(\sum_{i=1}^N \bar{\psi}_i\psi_i\right)^2-\left(\sum_{i=1}^N \bar{\psi}_i \gamma^5 \psi_i\right)^2\right] .
\end{equation}
This model has the $S$-matrix:
\begin{subequations}
\begin{align}
t_1(\hat \theta)&=f(\hat \theta,-\lambda),& t_2(\hat \theta)&=\frac{-\lambda}{1-\hat \theta}t_1(\hat \theta)\\
u_1(\hat \theta)&=t_1(1-\hat \theta),& u_2(\hat \theta)&=-\frac{\lambda}{\hat \theta}u_1(\hat \theta)\\
r_1(\hat \theta)&=0,& r_2(\hat \theta)&=0
\end{align}
\end{subequations}
where $\lambda=\frac{2}{N}$. This is the same as the original Class II minimal solution up to a CDD factor:
\begin{equation}
\frac{f(\hat \theta,-\lambda)}{f(\hat \theta,\lambda)}= \cos \left(\frac{\pi }{N}-\frac{\pi  \hat \theta}{2}\right) \sec \left(\frac{\pi
   }{N}+\frac{\pi  \hat \theta}{2}\right)
\end{equation}
The original Class II minimal solution has no poles on the physical sheet, but the $SU(N)$ chiral Gross-Neveu model does.  Physically this corresponds to a bound state in the antisymmetric tensor channel. The location of the pole is:
\begin{equation}
\theta=\frac{2i \pi}{N}.
\end{equation}
This is the only  bound state appearing in particle-particle scattering\footnote{Curiously, the antiparticle can be thought of as an $N-1$-particle bound state\cite{koberle1979scattering},\cite{kurak1979antiparticles}}. In particle-antiparticle scattering there are no $s$-channel poles, i.e. no bound states in the singlet or adjoint representations. 
 
We have four classes left. The Class III is identified with the $O(2N)$ $S$-matrix, which has already been discussed in the previous section. To the best of our knowledge it is not known to what field theories Classes IV, V, VI correspond to, and we will not attempt to reconstruct them in this work.

\section{Analytic bounds and properties}\label{sec:pole}
We now turn our attention to what possible exact statements we can make on the properties of $S$-matrices satisfying our assumptions of crossing symmetry, unitarity and analyticity. These assumptions imply constraints on possible couplings to bound states, which we will analyse numerically in the next section. Here we discuss how such bounds come about and how in some cases it is possible to derive optimal bounds analytically. In this section we set the mass of the external particle $m=1$.

Before we begin, it is convenient to introduce a new kinematic variable, $y$, defined by
\bea
s=\frac{2(1+y)^2}{1+y^2},\qquad y=\frac{2-\sqrt{s(4-s)}}{s-2}.
\eea
\begin{figure}[htpb]
\centering
\begin{minipage}[c]{0.5\textwidth}
\centering
\includegraphics[width=0.8\textwidth]{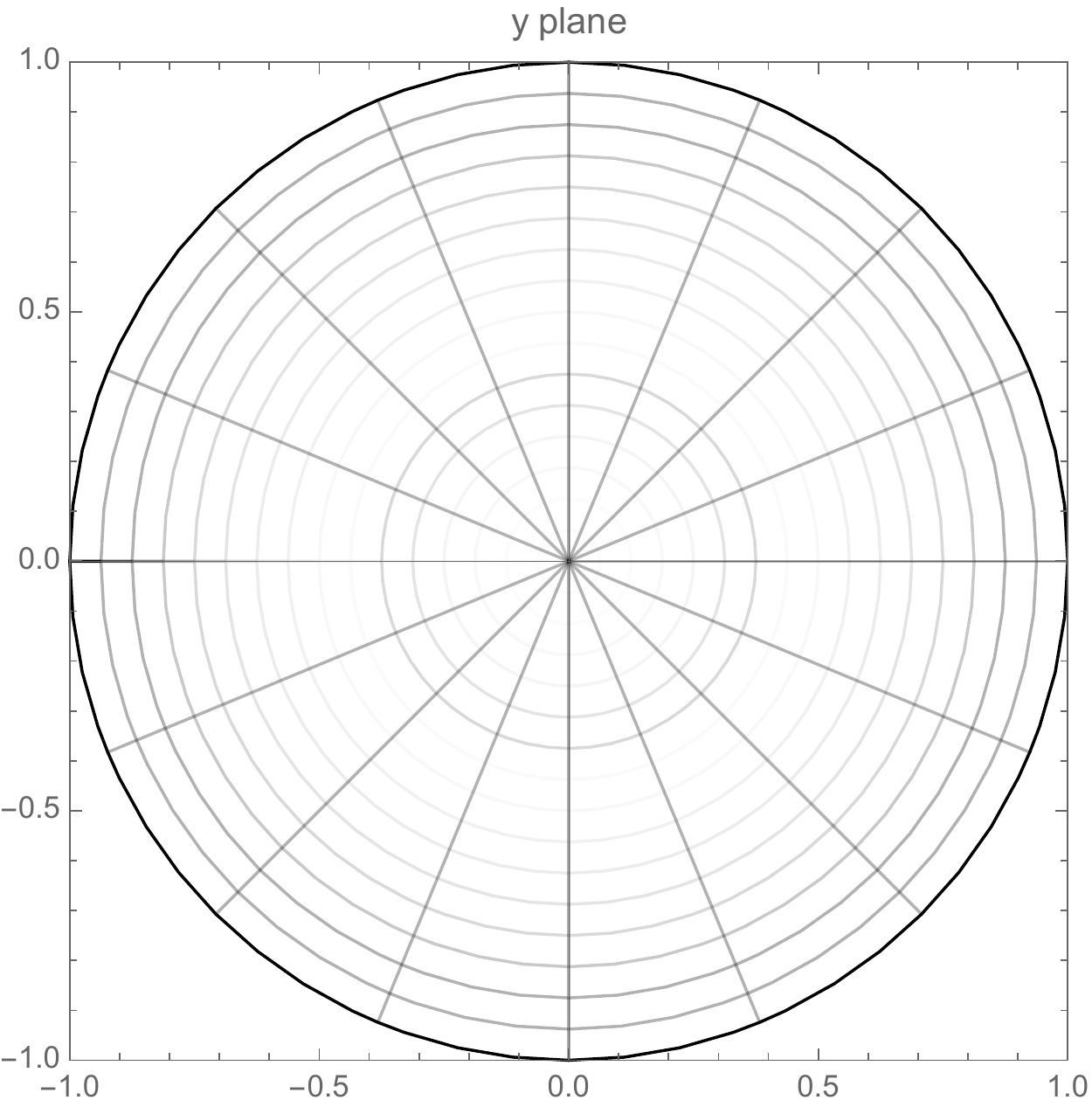}
\end{minipage}%
\begin{minipage}[c]{0.5\textwidth}
\centering
\includegraphics[width=0.8\textwidth]{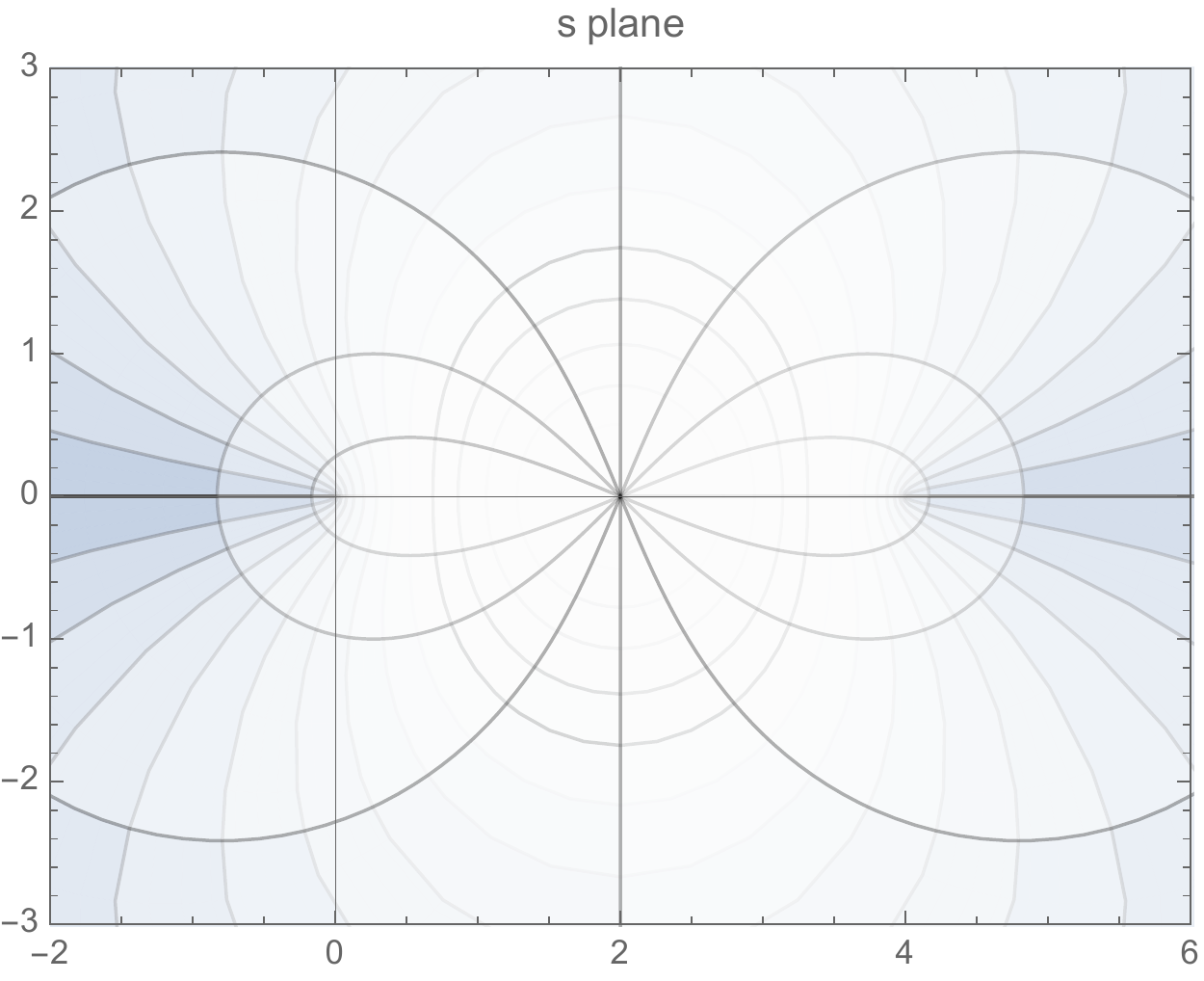}
\end{minipage}
\caption{$y$ variable}
\end{figure}
This conformal mapping transforms the $s$-plane excluding the cuts on $(-\infty,0)\cup (4,\infty)$ to the open unit disk $D=\{y: |y|<1\}$. The physical scattering region $s-i\epsilon>4$ corresponds to
\bea
P=\{(1-\epsilon)e^{i\phi}: \phi \in (0,\pi/2)\}
\eea
for sufficiently small $\epsilon>0$.  In what follows we will abuse notation and set $f(y)\equiv f(s(y)) \equiv f(\theta(y))$.

In general the problem we are interested in is to constrain the behaviour of a set of meromorphic functions on the disk, namely $S$-matrices $S^{(i)}(y)$ in various physical channels. The $S$-matrices obey the reality condition $\overline{S(y)}=S(\overline{y})$ and satisfy\footnote{The crossing conditions and ensuing results can be generalized easily to the $U(N)$ case.}
\bea
S^{(i)}(-y)=\sum_{j}\mathcal F^{(j,i)} S^{(j)}(y)\qquad \mbox{and}\qquad  |S^{(i)}(y)|\leq 1, \quad y\in P \cup \overline P,
\eea
i.e. crossing symmetry and unitarity respectively. Combining these two conditions together with the reality property, implies these functions should be bounded on the entire boundary of the disk:
\bea
|S^{(i)}(y)|\leq M\equiv \mbox{max}\left\{1,\sum_{j} |\mathcal F^{(j,i)}|\right\}, \qquad \forall y \in \partial D
\eea
This immediately implies that residues of poles of the $S$-matrices are bounded in modulus. The argument is straightforward \cite{Creutz1972a,Paulos2017}. Suppose some function $S^{(i)}(y)$ has $n$ poles at positions $p_k$ with residues $-r_k$, and define
\bea
g(y)\equiv S^{(i)}(y) \prod_{k=1}^n \frac{y-p_k}{1-y p_k}.
\eea
It follows that $g(y)$ is analytic on the disk and bounded on $\partial D$. By the maximal modulus principle, it must also be bounded on the entire disk, and we find:
\bea
|r_l|\leq \left |M\,(1-p_k^2)\prod_{k \neq l}^n \frac{p_l-p_k}{1-p_l p_k}\right|.
\eea
This in general not an optimal bound, since it does not take into account the full set of unitarity constraints on the unit disk, as well as those constraints following from other $S$-matrices. In the next section we will derive optimal bounds numerically in several circumstances.

As an aside, we should note that the physical sign of an $s$-channel pole residue is determined by the parity of the associated bound state~\cite{KAROWSKI1979244}. In general a given function $S^{(i)}$ will contain a proliferation of poles, both the physical $s$-channel as well as $t$-channel poles that follow from crossing symmetry. Incidentally, we note the connection between the residue in the $y$ variable and the physical coupling appearing in the scattering amplitude, viz.:
\bea
S(s)\sim -\frac{g_b^2 \mathcal J_b}{s-m_b^2}, \qquad \mathcal J_b=\frac{1}{2m_b \sqrt{4-m_b^2}},
\eea
then
\bea
r_b=g_b^2\, \frac{m_b\sqrt{4-m_b^2}-2}{\sqrt{m_b^2(4-m_b^2)}(m_b^2-2)^2}.
\eea
\subsubsection*{A special case}
There are a few special cases where it is possible to find exact $S$-matrices which saturate bounds on couplings although it is seems very difficult to find general solutions  as it was for the case without global symmetries \cite{Paulos2017}. Firstly, as a trivial case it is clear that by simply setting all $S$-matrices to be equal, and in particular individually crossing invariant, one recovers the problem without global symmetries. This follows essentially from the fact that the $\sum_{\mathcal P_i} T^{(i)}_{\delta \gamma,\alpha \beta}=\delta_{\alpha \gamma} \delta_{\beta \delta}$ and can be checked explicitly. So all such cases reduce to the problem considered in \cite{Creutz1972a,Paulos2017}. 

As a slightly less trivial example, consider the $O(2)$ $S$-matrix with a single bound state in the symmetric traceless, i.e. $(t)$, representation. Using crossing this leads to the following parameterization of the $S$-matrix:
\begin{subequations}
\label{o2example}
\bea
S^{(s)}(y)&=& \frac{r}{y+p}+N f(y)+g(y^2)+f(-y), \\
S^{(a)}(y)&=& \frac{r}{y+p}+g(y^2)-f(-y),  \\
S^{(t)}(y)&=& -\frac{r}{y-p}+g(y^2)+f(-y), 
\eea
\end{subequations}
We now note that $1=\sum_{i\in\{s,t,a\}} \mathcal F^{(t,i)}$ for all $N$. Hence $|S^{(t)}(y)|\leq 1$ on the entire disk and it is natural to try
\bea
S^{(t)}(y)=-\frac{1-p y}{y-p},
\eea
namely that $S$-matrix which saturates the bound on the coupling $r$ discussed previously. If we can find $S^{(s)}(s), S^{(a)}$ partners that satisfy crossing and unitarity, we will have shown that this bound is optimal. It is easy to check that
setting $r=1-p^2, g(y^2)=p$ and $f(y)=0$ in the equations above does the job. Hence $|r|=1-p^2$ is the optimal bound.

\subsubsection*{Reflectionless property for $SU(N)$ Gross-Neveu}
It is possible to play similar kinds of games to find other special solutions, but we will not do this here. Rather we now discuss a particularly useful property for our numerical setup, which relates to the fact that the reflection amplitude for the $SU(N)$ Gross-Neveu model vanishes, i.e.
\bea
S^{\mt{sing}_+}-S^{\mt{sing}_-}=S^{\mt{adj}_+}-S^{\mt{adj}_-}=0
\eea
or equivalently $r_1=r_2=0$ in the notation of equations \reef{eq:undecomp}. We would like to discuss under which assumptions this is the case.
Consider the unitarity constraints which hold in the physical region,
\begin{equation}\label{cons1}
|u_1\pm u_2|\leq 1
\end{equation}
\begin{equation}\label{cons2}
|t_1\pm r_1|\leq 1
\end{equation}
\begin{equation}\label{cons3}
|t_1\pm r_1+N(t_2\pm r_2)|\leq 1,
\end{equation}
and suppose we are maximizing the residue $g_1$ of a particular pole that does not appear in $r_1,r_2$. We also assume that the overall pole structure is fixed in such a way that possible poles appearing in $r_1,r_2$ do not appear elsewhere. We denote \textit{problem} $1$ this optimization problem, which must satisfy the constraints above together with the crossing equations \reef{uncross} which we repeat here:
\bea
u_1(\theta)=t_1(i\pi-\theta)\qquad 
u_2(\theta)=t_2(i\pi-\theta) \qquad 
r_1(\theta)=r_2(i\pi-\theta).
\eea
If we further impose the constraint $r_1=r_2=0$, as well as all the constraints from \textit{problem} $1$, this maximization problem is called \textit{problem} $2$, and the corresponding maximal residue is $g_2$. 

We first note that $g_2$ must be smaller than $g_1$ because there are more constraints. But we can also get $g_1\leq g_2$. Indeed, from the first set of inequalities we can obtain
\begin{equation}
|u_1\pm u_2|\leq 1
\end{equation}
\begin{equation}
|t_1|^2+|r_1|^2\leq 1
\end{equation}
\begin{equation}
|t_1+N t_2|^2+|r_1+N r_2|^2\leq 1
\end{equation}
Since $r_1$ and $r_2$ are related by crossing, and none of its poles appear in other functions, they effectively form a decoupled subsector, and the equations above are stronger constraints on $t_1,t_2,u_1,u_2$ than those of problem 2, namely
\begin{equation}
|u_1\pm u_2|\leq 1
\end{equation}
\begin{equation}
|t_1|^2\leq 1
\end{equation}
\begin{equation}
|t_1+N t_2|^2\leq 1.
\end{equation}
 Overall then, $g_1\leq g_2\leq g_1 \Rightarrow g_1=g_2$, and hence it is consistent to set $r_1=r_2=0$.

Note that if $r_1$ and $r_2$ had extra poles appearing in the other functions, the argument would fail, since then these extra poles could shield the contribution of the one whose residue we are maximizing, and hence a higher coupling might have been obtained by keeping them non-zero. This expectation is borne out in concrete examples.

\section{Numerical results}
\label{sec:num}
In this section we present our results for determining upper bounds on couplings to bound states. But first, let us describe the general setup. The reader can keep in mind the special $O(2)$ case discussed in the previous section as an example. Firstly, one chooses a set of physical $s$-channel poles appearing in individual channels in a 2-to-2 scattering process: for instance, a bound state in the tensor channel together with another in the singlet sector in the $O(N)$ case.
 Using the crossing relations $S(i\pi-\theta) \sim \sum \mathcal F S(\theta)$ described in section 2, this implies the existence of other, cross-channel poles, which we must also include. Once this is done, the pole structure is fixed, and whatever remains must be analytic on the disk in the $y$ variable and in particular can be approximated by a polynomial. Schematically
\bea
S^{(i)}(y)\sim -\sum_{m} \frac{r_m^{(i)}}{y-p_m}-\sum_{n} \frac{\tilde r_n^{(i)}}{y-\tilde p_n}+\sum_{k=0}^{M} a_m^{(i)} y^m
\eea
where the first and second sums run over direct and cross channel poles respectively.
That is, in the example of the previous section, in equations \reef{o2example} one would approximate the functions $f,g$ by polynomials of finite degree $M$. Finally, we want maximize the residue of a particular pole while imposing the unitarity constraints in each channel. Recall the constraints hold for $y=e^{i\phi}$ with $\phi\in(0,\pi/2)$. In practice we check unitarity only on an evenly spaced grid of $K$ points in this region. To find the maximum residue we use Wolfram {\tt Mathematica}'s function \texttt{FindMaximum}.
Generally, the numerical bound increases as $M$ goes up, but as we do this we must use higher $K$ to ensure that the unitarity condition is being correctly taken into account. Because $M$ is proportional to the highest frequency on the boundary of the unit circle, we should choose $K\propto M$. Empiricaly setting $K=2M$ is enough to ensure the unitary condition. Finally we increase $M$ until the maximum residue is varying negligibly.

\subsection{$O(N)$ bounds}

We begin by considering $S$-matrices with global $O(N)$ symmetry. We present bounds for $S$-matrices with the same bound state spectrum as known integrable models, i.e. the Sine-Gordon Model and $O(N)$ Gross-Neveu model. More general bound state spectra are considered in Appendix \ref{On} -- although we know no integrable models which saturate such bounds, they still hold for general $1+1$ gapped QFTs. 

\subsubsection{$N=2$, comparison with Sine-Gordon model.}

\begin{figure}[htbp]
\centering
\includegraphics[width=\textwidth]{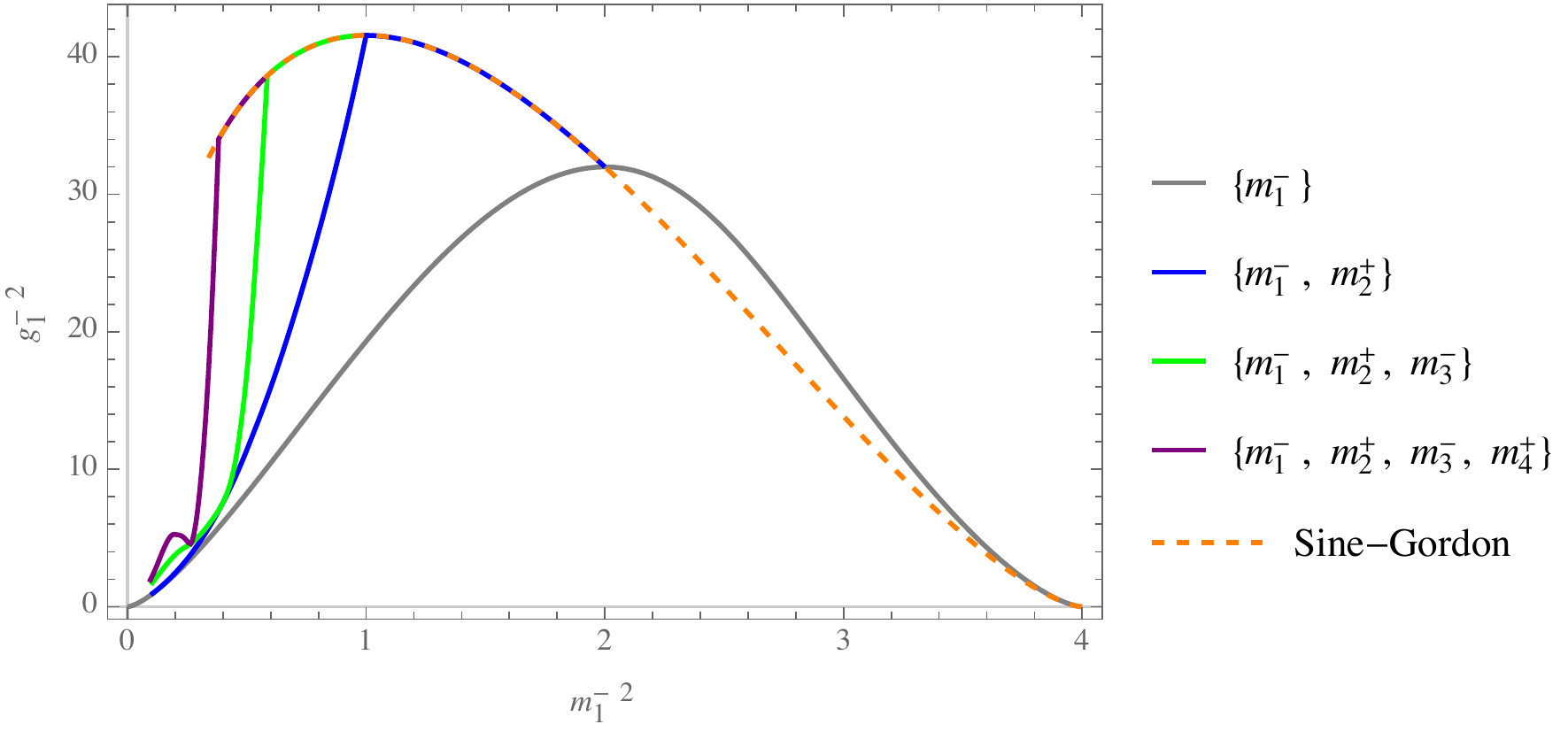}
\caption{\label{fig:SGcom} Bound of a coupling to lightest pseudoscalar particle in the presence of other scalar or pseudoscalar states, and comparison with the sine-Gordon model, which saturates the bound as soon as two particles are present. Further explanations in the main text. The x-axis is the pole of the lightest bound state, and the y-axis is the corresponding residue bounds. We find there are is a sharp turn at the point when new bound states are created in the Sine-Gordon model.}
\end{figure}

In fig. \ref{fig:SGcom} we present our numerical bounds for $S$-matrices with the same pole structure as the sine-Gordon model $S$-matrix that describes soliton scattering. In particular we consider an upper bound on the residue of the lightest (pseudoscalar) particle, which lives in the antisymmetric representation of $O(N$). The lightest bound state becomes lighter and lighter as the renormalized coupling constant $\xi$ decreases. At the same time, new bound states appear from the multiparticle region. We include these at appropriate points, with the same mass-ratios as for the sine-Gordon model. 

\begin{figure}[htbp]
\centering
\includegraphics[width=0.9\textwidth]{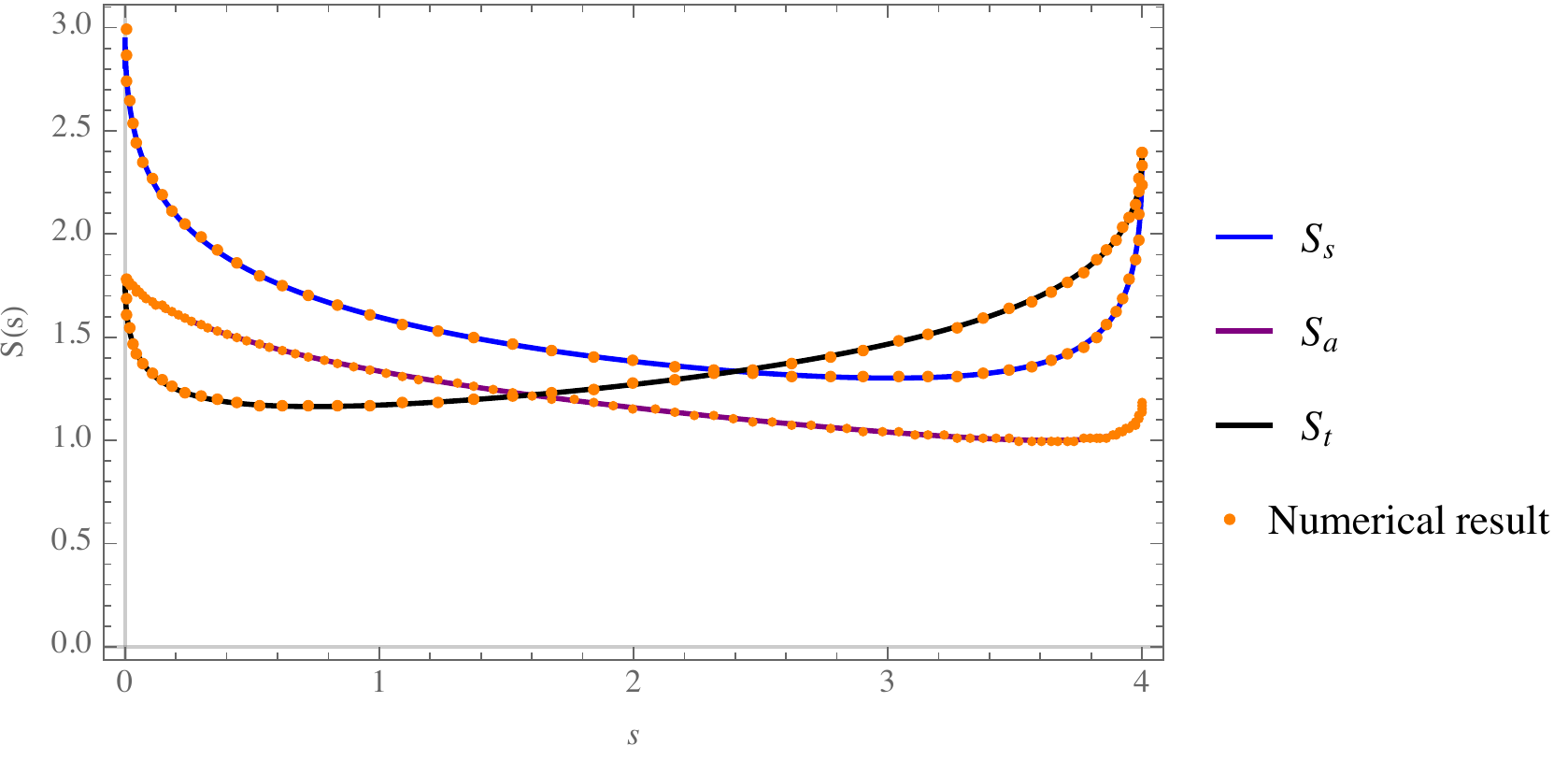}
\caption{\label{fig:smatrix}Comparision of the numerical $S$-matrix with the analytical sine-Gordon matrix with $\frac{\pi}{\xi}=2.5$. We have subtracted the pole structure in $S(s)$ for a cleaner plot.}
\end{figure}

\begin{figure}[htbp]
\centering
\includegraphics[width=\textwidth]{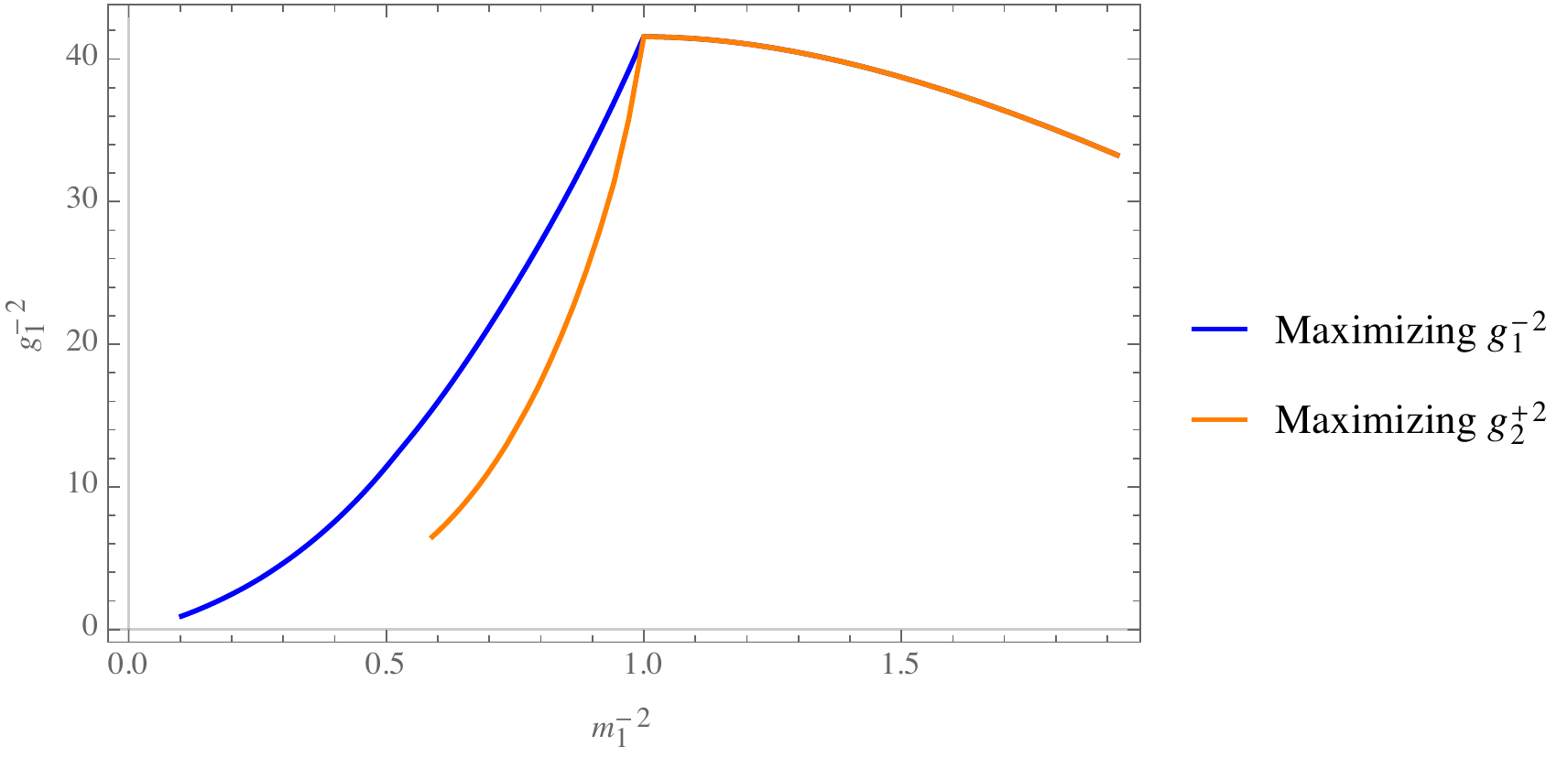}
\caption{\label{fig:SGdif}Comparison of bounds when we maximize the pseudoscalar or scalar couplings. The orange line is the curve for the pseudoscalar coupling when we maximize the scalar coupling. The blue line is the curve for the pseudoscalar coupling when we maximize the pseudoscalar coupling. By this figure we can verify Eq. \ref{constraint}}
\end{figure}

A few remarks regarding this plot:

\begin{itemize}
	\item We note that the $S$-matrix that maximizes the coupling of the lightest bound state is the same as the integrable model with the exception of the case where there is a single bound state, i.e. when $\pi/2<\xi\leq \pi$. This is a surprising result, considering the highly non-trivial nature of the relevant $S$-matrix. We can reproduce the integrable $S$-matrix from our numerics, as it is shown in fig. \ref{fig:smatrix} for the two mass case. 
	
	\item Concerning the maximal solution for the bound when there is only one pseudoscalar, we have checked that it is not a simple CDD extension of the minimal solution. In particular it has the same pole structure but without zeros (in contrast to the sine-Gordon model.) It is likely that adding information about the zero we could reproduce the sine-Gordon $S$-matrix, but we will not do this here.
	\item An interesting feature is that when an additional pole comes down from the multiparticle region to threshold, there is an abrupt change in the numerical bound. We can observe numerically that this occurs when two distinct poles coincide. Note that the coupling bound with more bound states is strictly higher than that with less bound states, as it should be. In fig. \ref{fig:SGcom}, we can see a bounce when two bounds coincide. We can further predict where these bounces are located: for the range of $\xi$ where there are $n$ bound states, the $n$th bound state will coincide with the first $n-1$ bound state's cross channel poles in order, and produce $n-1$ kinks on the bound. These kinks correspond to the coincidence of the $n$th bound state and $k$th bound state which occurs at:
	\begin{equation}
		(n+k)\xi=\pi
	\end{equation}
	This is consistent with the plot.
	\item One may wonder what would happen if instead of maximizing the lightest pseudoscalar coupling we made a different choice.  Will we get the same result? Or in other words, do all the couplings maximize at the same time? The answer is: when there is an integrable model located at the bound, all the couplings maximize at the same time. But in general, this is not guaranteed. In fig. \ref{fig:SGdif}, we see that when there is no corresponding integrable model, the couplings do not maximize simultaneously. Experimentally it seems that, at least for the $O(N)$ model with antisymmetric tensor and scalar bound states, the couplings are maximized simultaneously when
	\begin{equation}\label{constraint}
		m_2^2+m_1^2\geq 4.
	\end{equation}

	\item Finally, let us note that we observe numerically very good convergence of the bound value. Any given point converges easily and takes only $M=20$ (recall $M$ is the degree of the polynomial approximation) to get a very good match with the sine-Gordon model.
\end{itemize}

\subsubsection{\texorpdfstring{$N\geq 3$}-, comparison with Gross-Neveu model}

\begin{figure}[htbp]
\centering
\includegraphics[width=\textwidth]{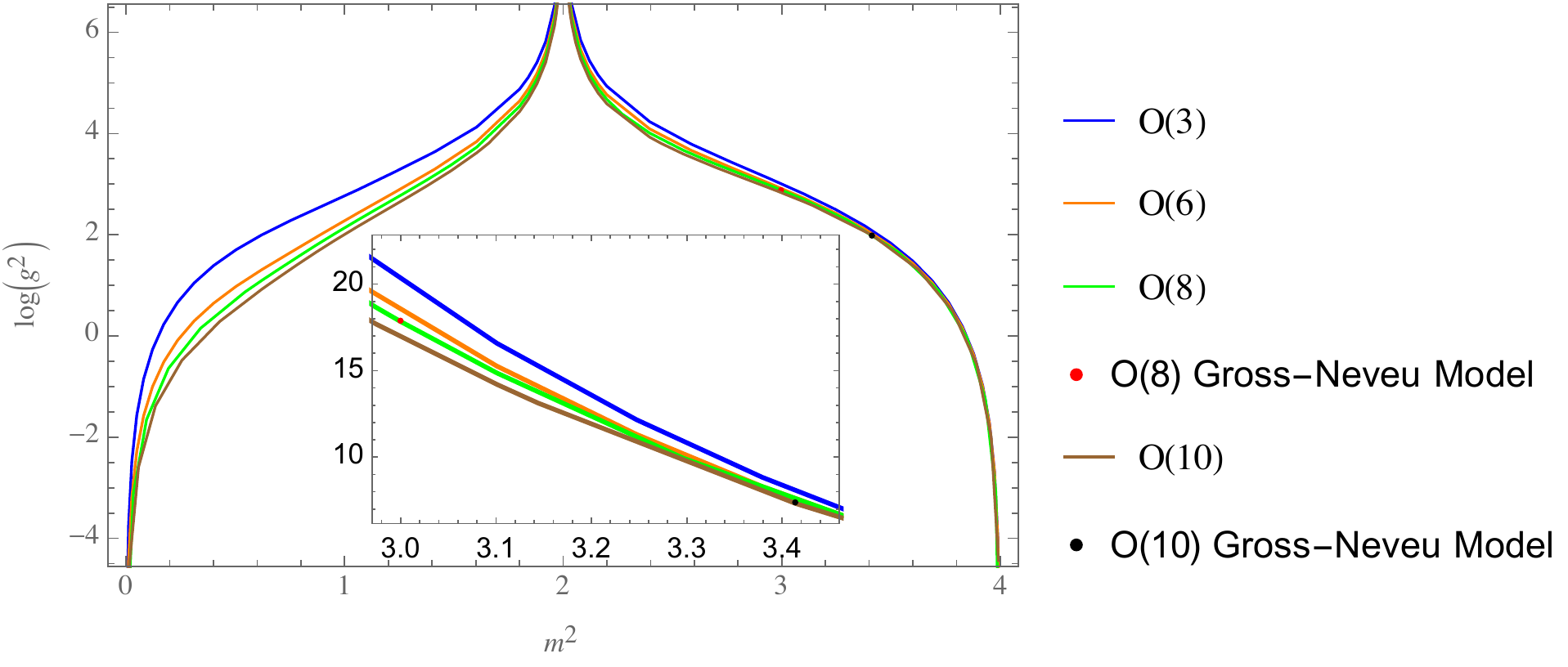}
\caption{\label{fig:GNOn}Numerical bounds and the Gross-Neveu model points. This is a bound for the coupling to an antisymmetric tensor in the presence of a scalar bound state of the same mass.}
\end{figure}

In fig. \ref{fig:GNOn} we show our numerical bounds and a comparison with the $O(N)$ Gross-Neveu model. This is the plot for a $O(N)$ S-matrix with a scalar and antisymmetric tensor bound states with degenerate masses, which is the bound state structure of the $O(N)$ Gross-Neveu model. 
In the plot, we consider $O(3)$, $O(6)$, $O(8)$ and $O(10)$. We note that only $O(8)$ and $O(10)$ Gross-Neveu model have the desired bound state structure -- $O(3)$ has no poles and $O(6)$ has second order poles.

\begin{figure}[htbp]
\centering
\includegraphics[width=0.9\textwidth]{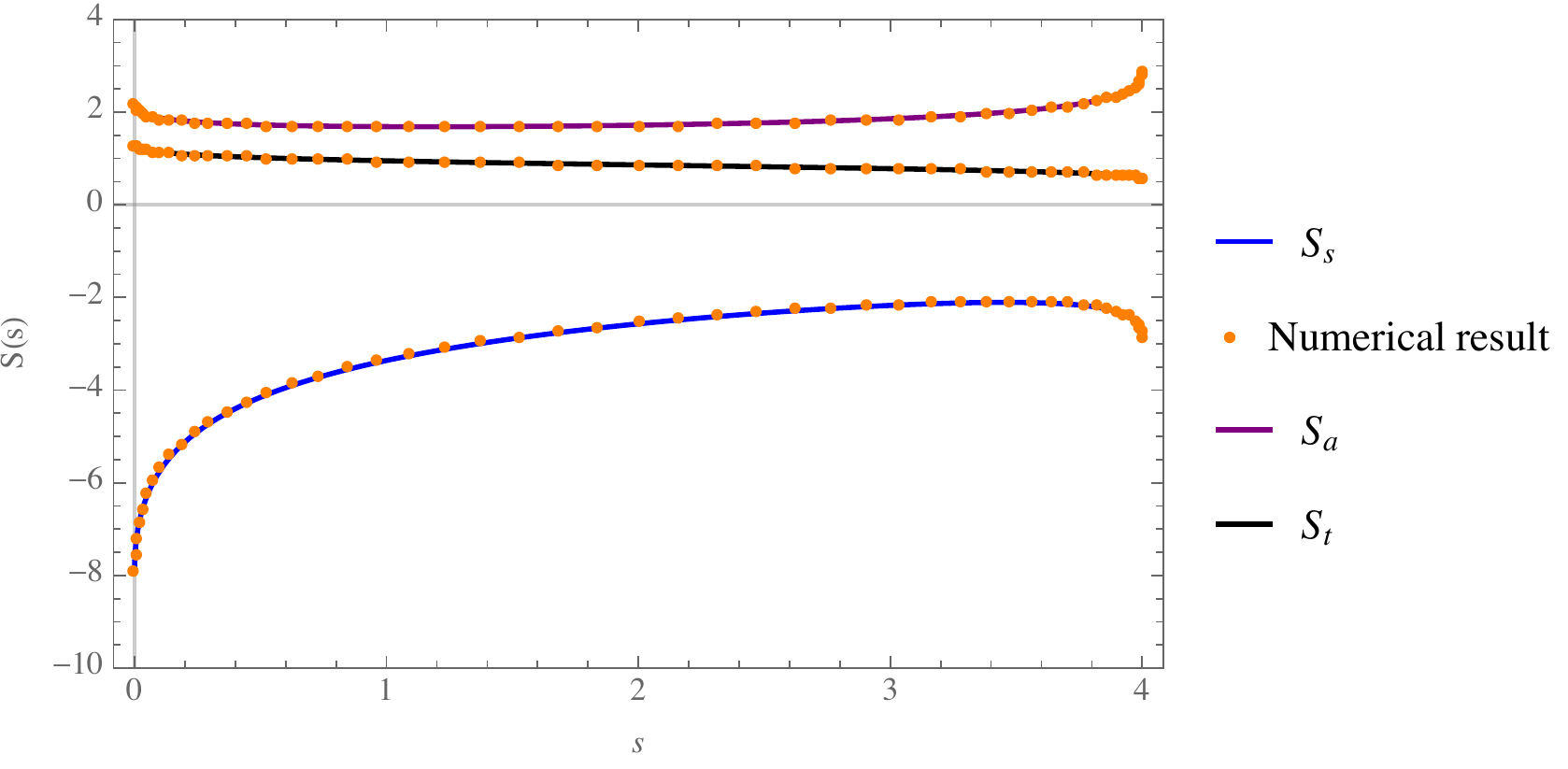}
\caption{\label{fig:O8smatrix}Comparison of the numerical $S$-matrix with the $O(8)$ $S$-matrix. We have subtracted the pole structure of each $S$-matrix individually for clarity.}
\end{figure}

Some remarks regarding this plot:

\begin{itemize}
	\item We note that the $S$-matrices for the $O(N)$ Gross-Neveu model saturate our bounds, except for $O(5)$. Although not shown, we have checked that this is true for $7\leq N\leq 11$. As shown in fig. \ref{fig:O8smatrix}, the numerical $S$-matrix at the appropriate value of the bound state mass is exactly the same as the corresponding integrable $S$-matrix. 
The $O(6)$ Gross-Neveu model has a double pole, and so should be thought of as being located at $(2,\infty)$ in this plot, so it is also consistent with the bound. 
	\item The $O(5)$ model does not lie on the bound, and one can check that our numerical optimal $S$-matrix is not a simple CDD extension of the minimal solution. Part of the issue is that $O(5)$ has the wrong sign for the $s$-channel residue, but multiplying the whole $S$-matrix by a sign does not fix this problem. It is likely that adding further constraints might be able to lower the bound enough for a match. 
	\item We observe numerically that convergence in this case is not as good, especially on the left hand side of the bound curve, requiring high degree polynomial approximations~($M \sim 40$). 
\end{itemize}

\subsection{$U(N)$ bounds}
We now apply the same method as in the last section to obtain numerical results for the $U(N)$ model. 

\subsubsection{Comparison with $SU(N)$ Gross-Neveu model}
\begin{figure}[htbp]
\centering
\includegraphics[width=0.9\textwidth]{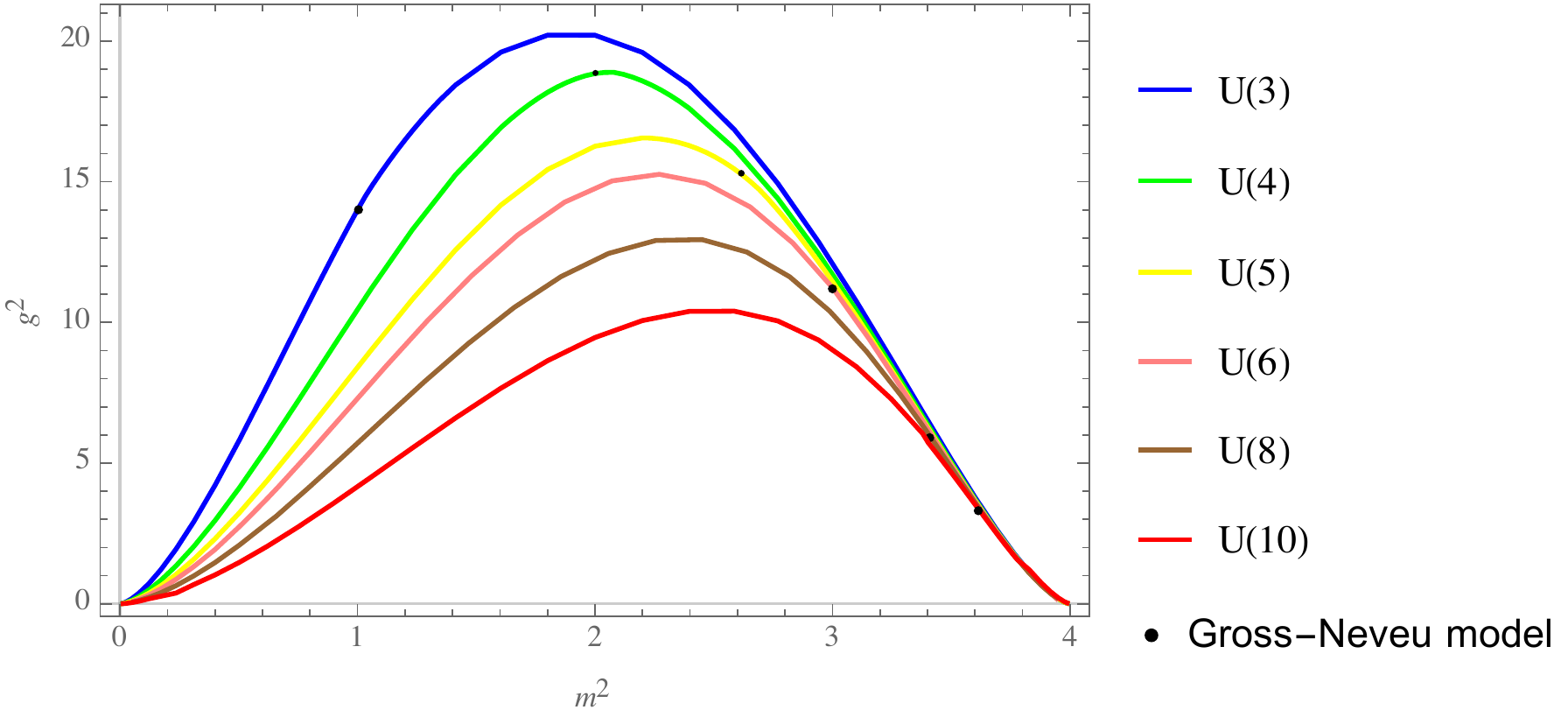}
\caption{\label{fig:Uncompn}Numerical bounds for the coupling to a single bound state in the antisymmetric tensor channel. The Gross-Neveu model saturates the bound for all $N$ we consider.}
\end{figure}

\begin{figure}[htbp]
\centering
\includegraphics[width=0.9\textwidth]{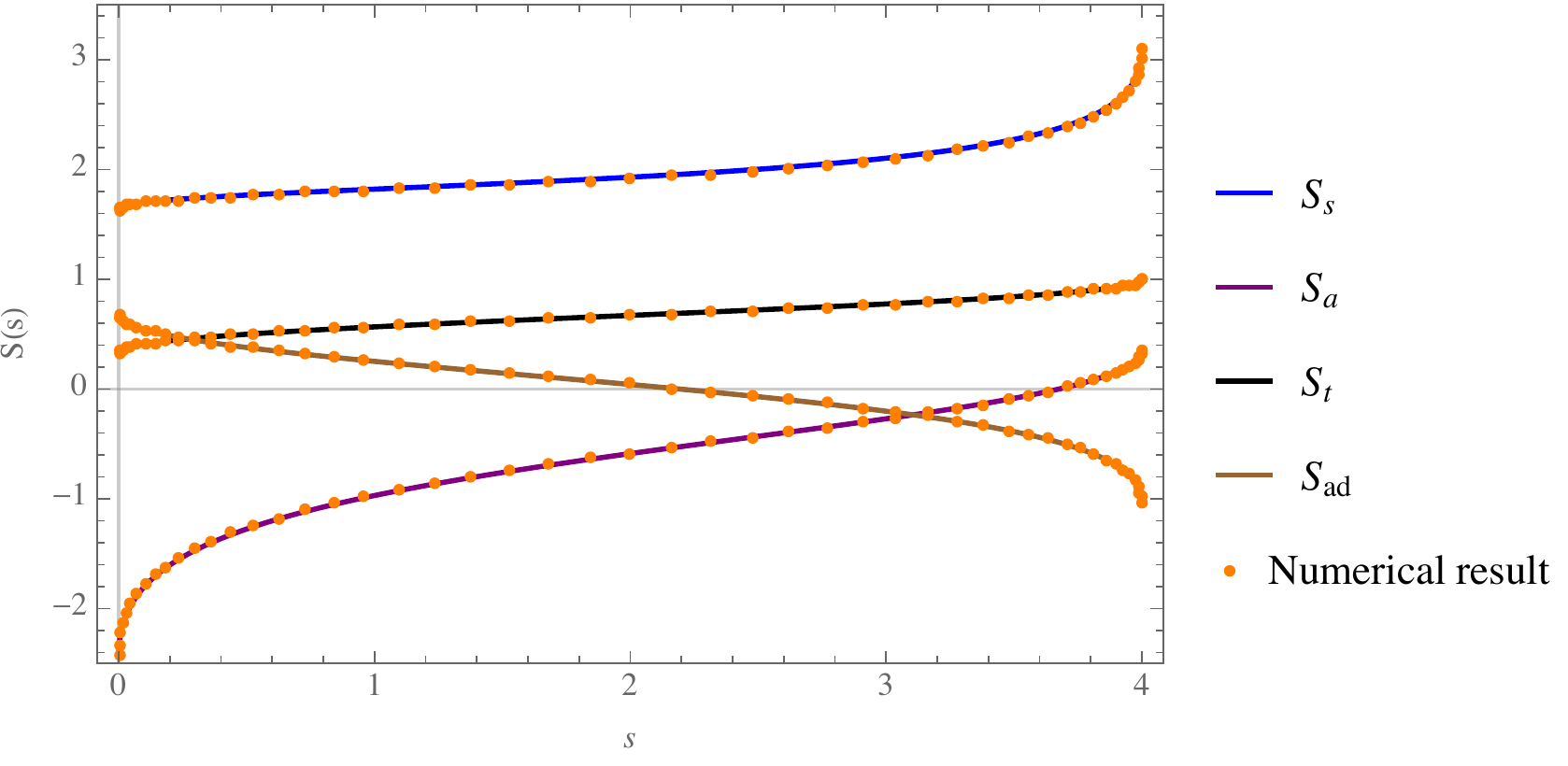}
\caption{\label{fig:U3smatrix}Comparison of the numerical $S$-matrix with the $SU(3)$ Gross-Neveu model $S$-matrix. We have subtracted the pole structure in $S(s)$.}
\end{figure}

As we have seen in section 3, the $SU(N)$ Gross-Neveu model only has an antisymmetric tensor bound state. We maximize the corresponding coupling for several values of $N$ to obtain figure \ref{fig:Uncompn}. We find that the $SU(N)$ Gross-Neveu model lies exactly on the coupling bound. We can also find the $S$-matrix for $SU(N)$ Gross-Neveu model by maximizing the bound at correct value of the bound state mass. The comparison with the analytical $S$-matrix for the $SU(3)$ Gross-Neveu Model is shown in figure \ref{fig:U3smatrix}.

\subsubsection{General bounds}

\begin{figure}[htbp]
\centering
\includegraphics[width=0.9\textwidth]{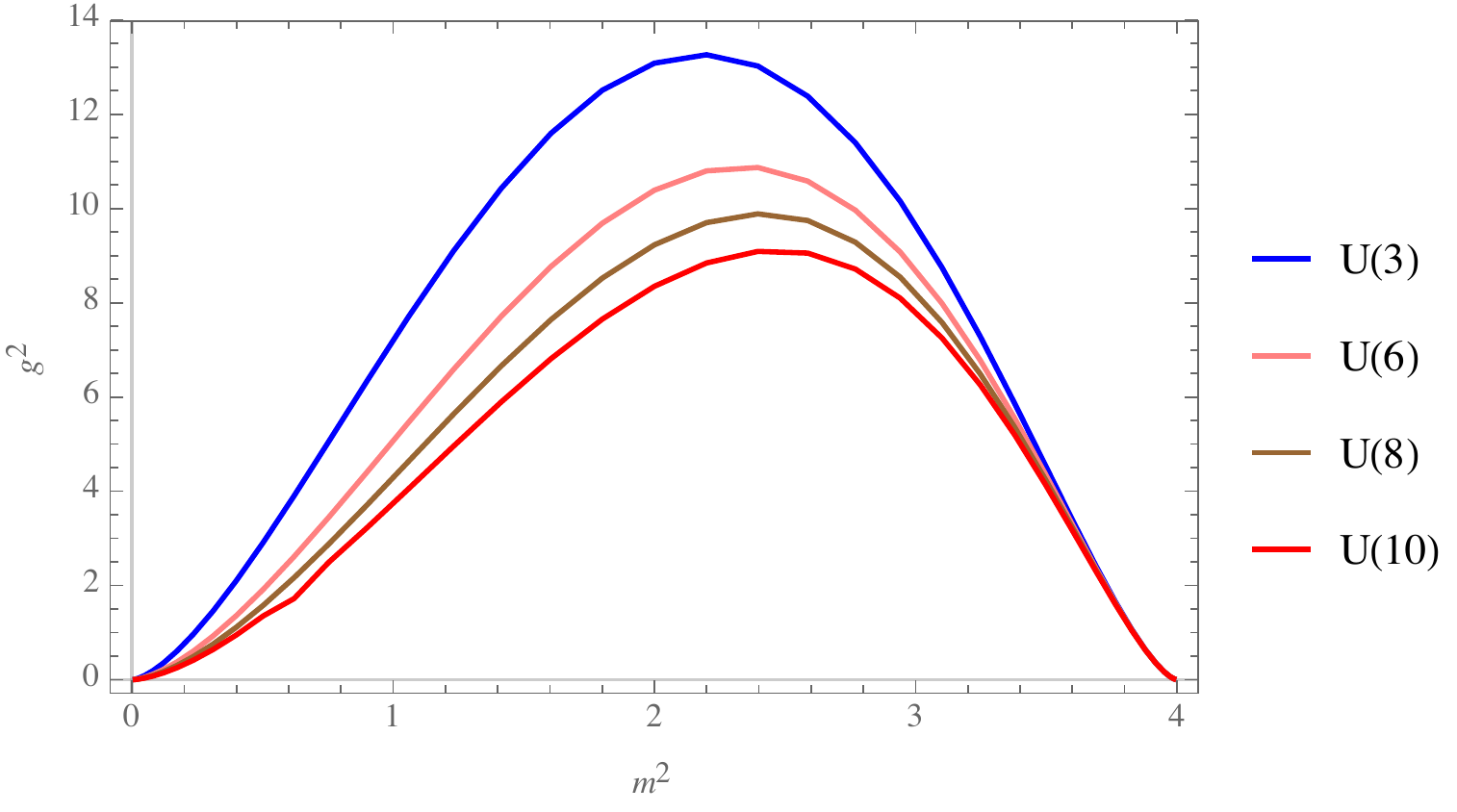}
\caption{\label{fig:Unsym}Upper bounds on the coupling to a single symmetric tensor bound state.}
\end{figure}

Here we consider the coupling bound for the case where there is only one bound state under $U(N)$ symmetry. Theoretically, there should be six plots corresponding to six channels, but the bounds for $\mathrm{singlet}^+$ and $\mathrm{singlet}^-$ are the same, as they are for $\mathrm{adjoint}^+$ and $\mathrm{adjoint}^-$. Since we have already discussed the antisymmetric tensor case above, there are three further channels shown below. Figures \ref{fig:Unsin}, \ref{fig:Unadjoint} and \ref{fig:Unsym}  correspond to bounds on the singlet, adjoint and symmetric tensor channels respectively.

\begin{figure}[htbp]
\centering
\includegraphics[width=0.9\textwidth]{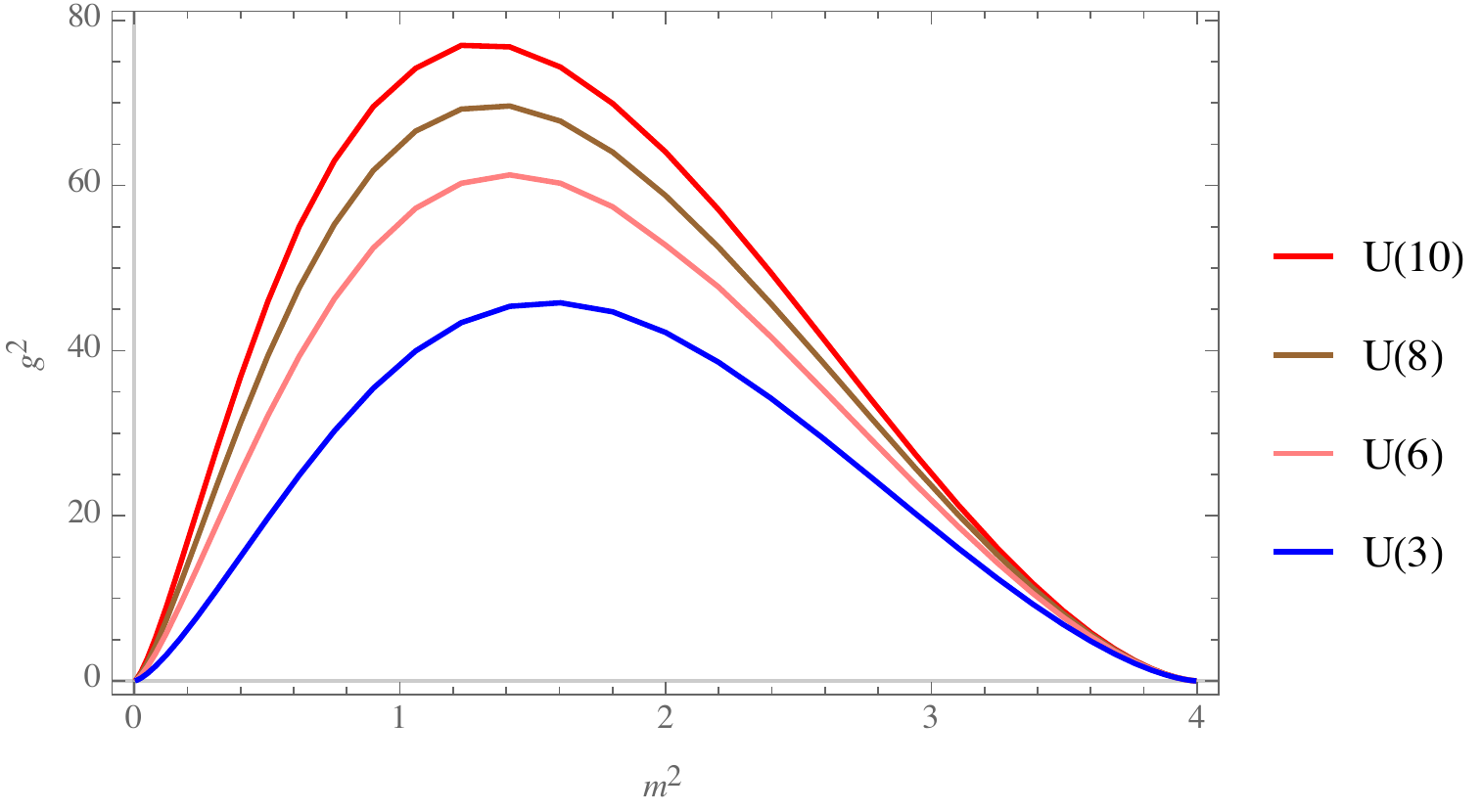}
\caption{\label{fig:Unsin}Upper bound on the coupling to a bound state in the singlet channel in the presence of no others. The bound is increasing with $N$ but $g^2/N$ is decreasing as it should.}
\end{figure}

\begin{figure}[htbp]
\centering
\includegraphics[width=0.9\textwidth]{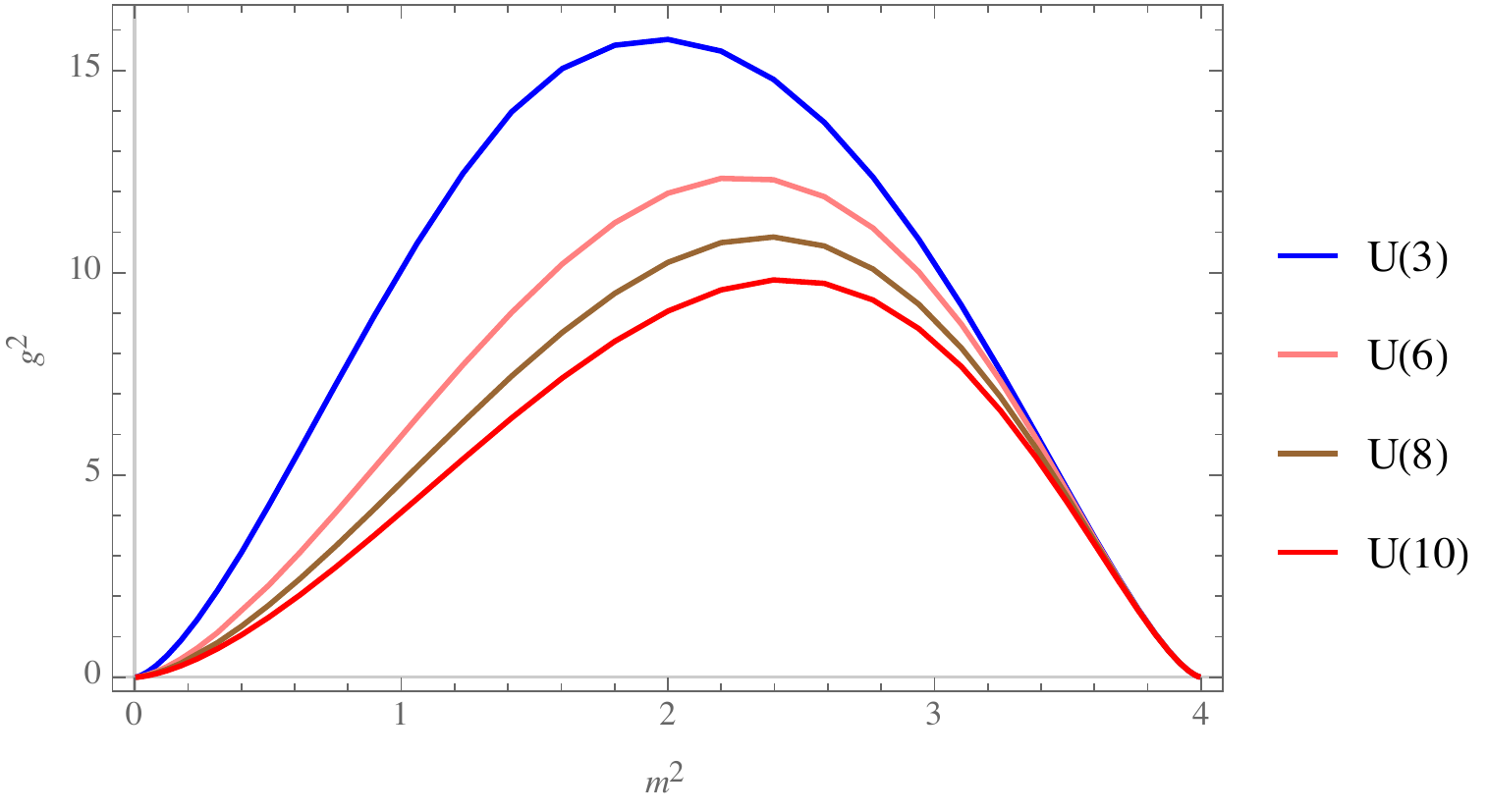}
\caption{\label{fig:Unadjoint}Upper bounds on the coupling to a single adjoint bound state.}
\end{figure}

 \section*{Acknowledgements} 
M.F. Paulos thanks the organizers of the Simons Non-perturbative Bootstrap workshop on the $S$-matrix Bootstrap in the Azores for providing a very stimulating environment while this work was being completed. The authors thank also L.G. C\'ordova and P. Vieira for discussing with us their related upcoming work. Z. Zheng is supported by an International Selection Scholarship from the \'Ecole Normale Sup\'erieure, Paris, France.

\appendix 

\clearpage
\section{Further results for $O(N)$}\label{On}

Here we show extra bounds for the $O(N)$ case. In particular we consider coupling bounds in simple cases containing only one or two exchanged states in various channels. Recall that $(s), (t), (a)$ correspond to the scalar, symmetric traceless tensor and antisymmetric tensor representations of $O(N)$, respectively.

\subsection{Single bound state}
\begin{figure}[htbp]
\centering
\includegraphics[width=0.9\textwidth]{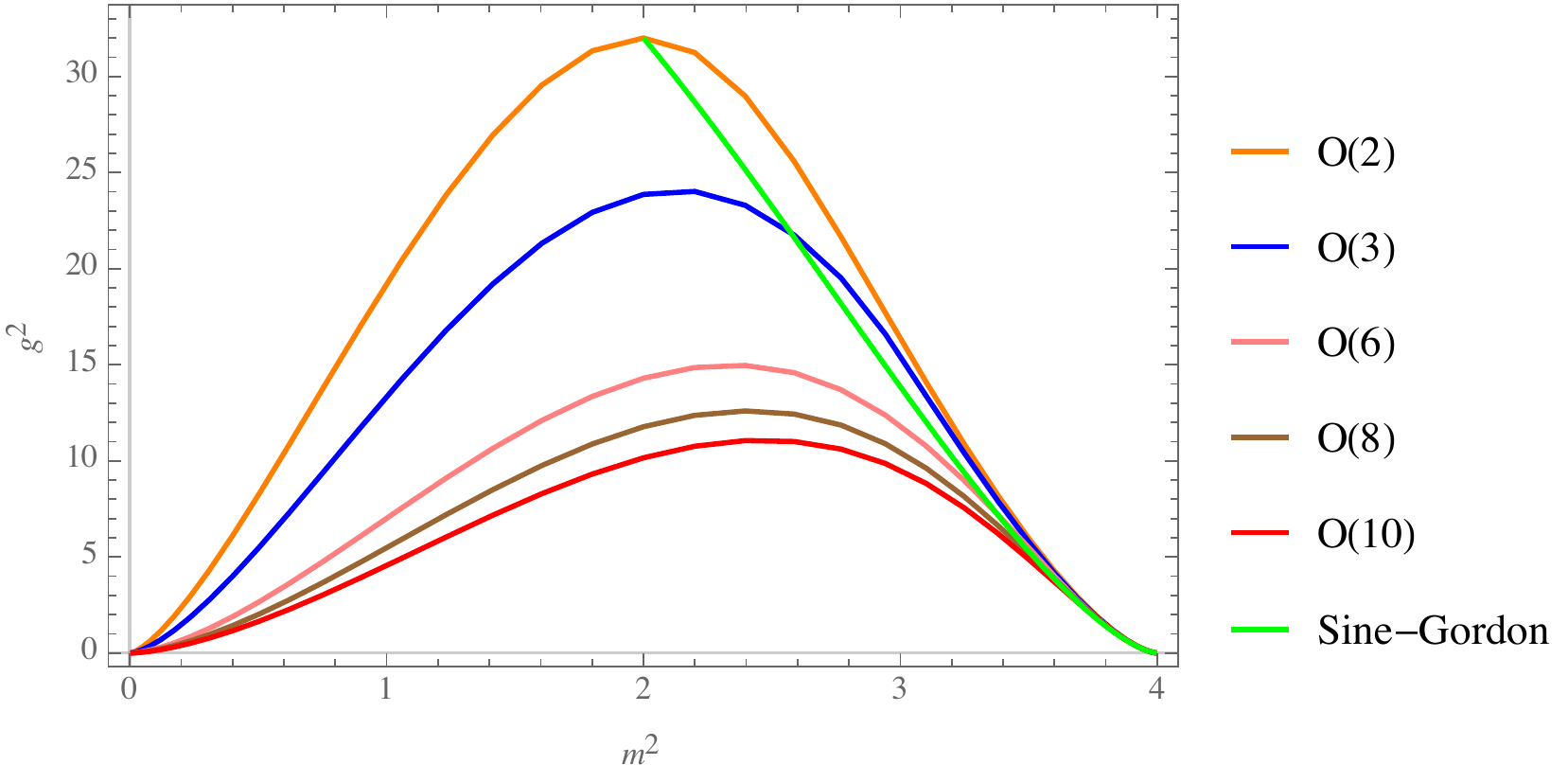}
\caption{\label{fig:Onanti} Coupling bound when there is only one antisymmetric tensor(pseudoscalar for $O(2)$ case). For comparison we also show the exact coupling constant for the sine-Gordon model with a single pseudoscalar, i.e. when $\xi>\pi$.}
\end{figure}

\begin{figure}[htbp]
\centering
\includegraphics[width=0.9\textwidth]{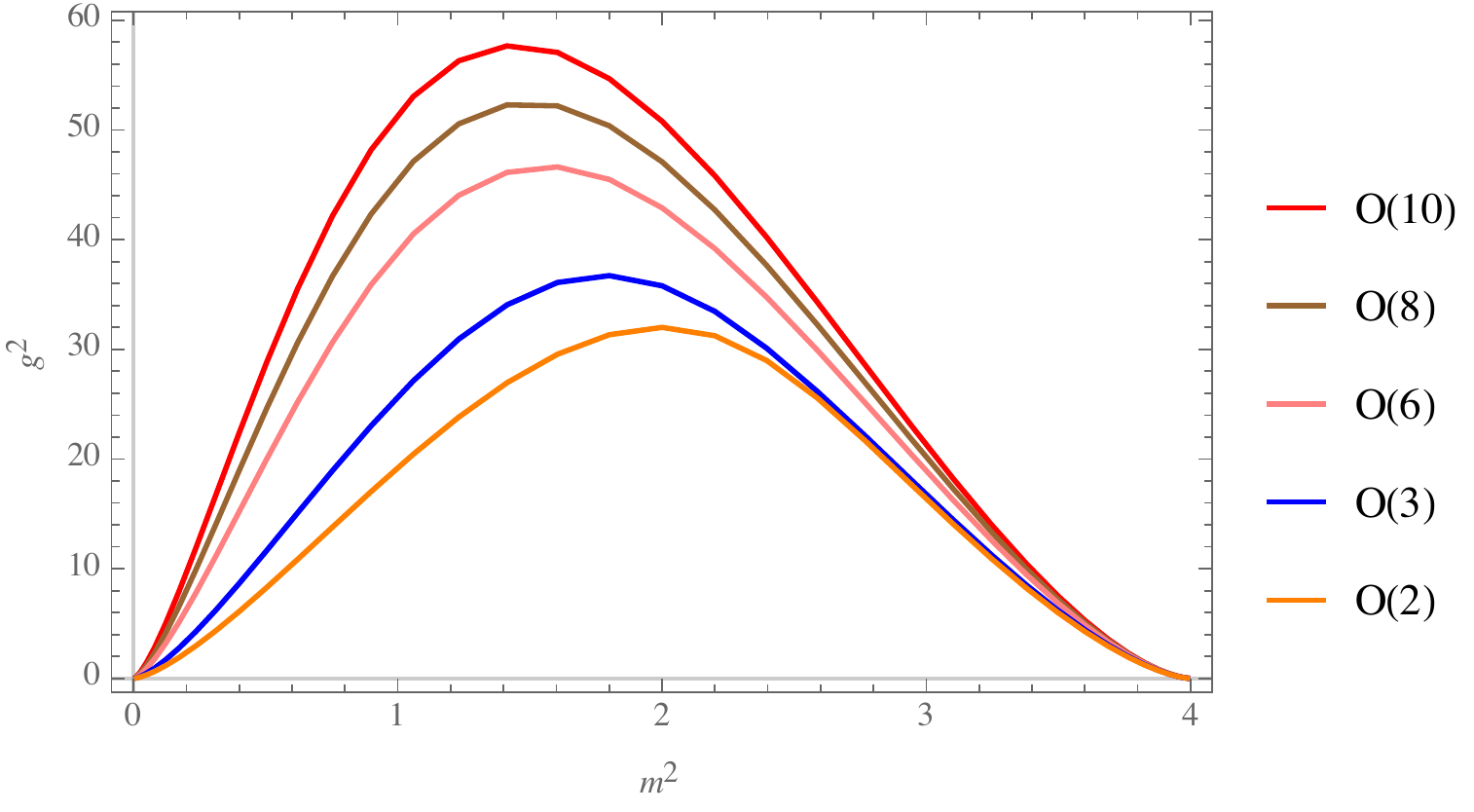}
\caption{\label{fig:Onscalar}Coupling bound to an $(s)$ bound state. Although the bound increases with $N$, the appropriately normalized $g^2/N$ decreases.}

\end{figure}

\begin{figure}[htbp]
\centering
\includegraphics[width=0.9\textwidth]{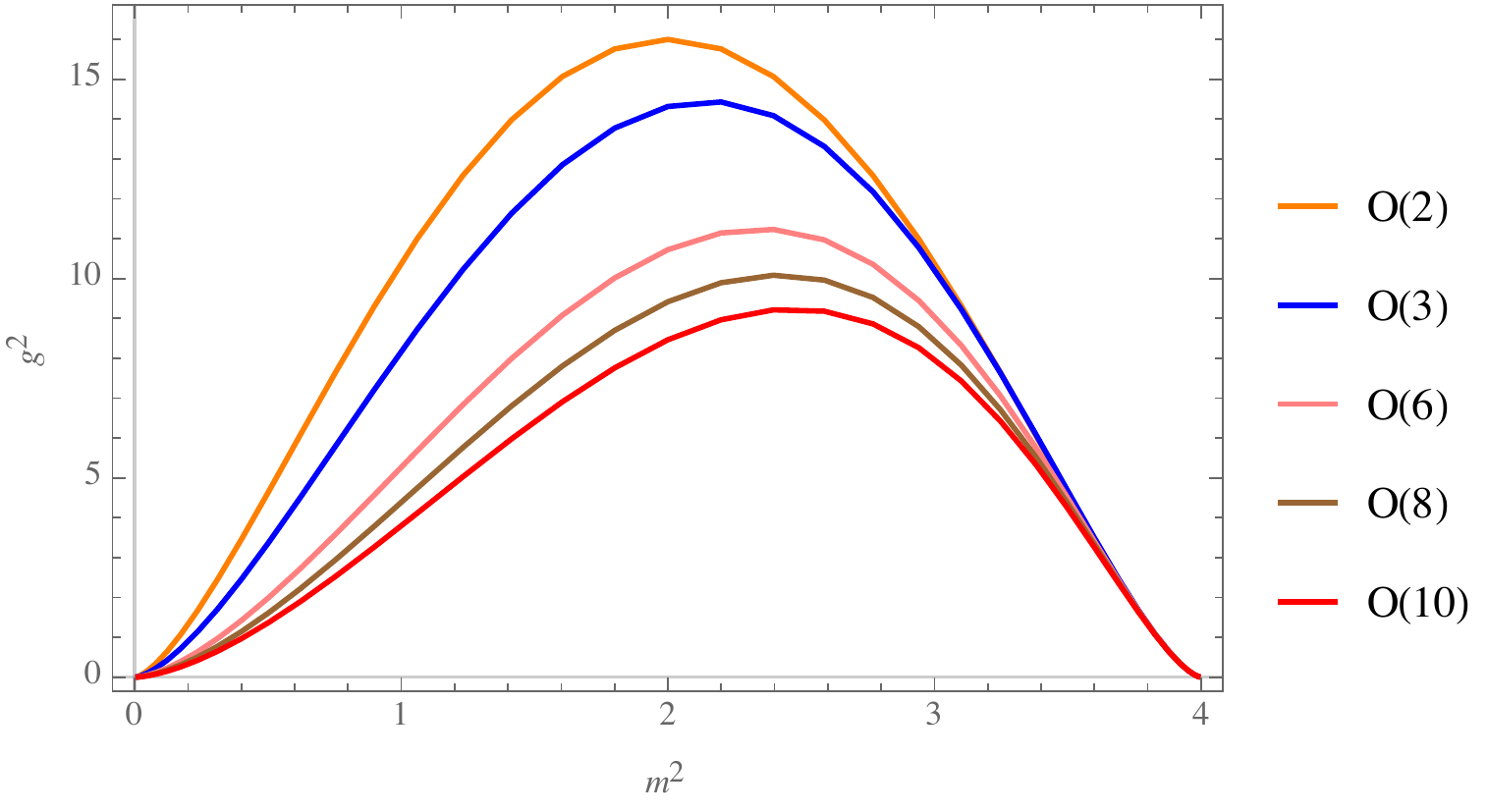}
\caption{\label{fig:Onsym}Coupling bound to a $(t)$ state with no extra bound states. }
\end{figure}

In fig. \ref{fig:Onanti} we show bounds for the coupling to an antisymmetric tensor particle (pseudoscalar for $N=2$). Figures \ref{fig:Onscalar} and \ref{fig:Onsym} repeat the analysis for a bound state in the $(s)$ and $(t)$ representations, respectively.

\clearpage

\subsection{Two bound states}
\begin{figure}[htbp]
\centering
\includegraphics[width=\textwidth]{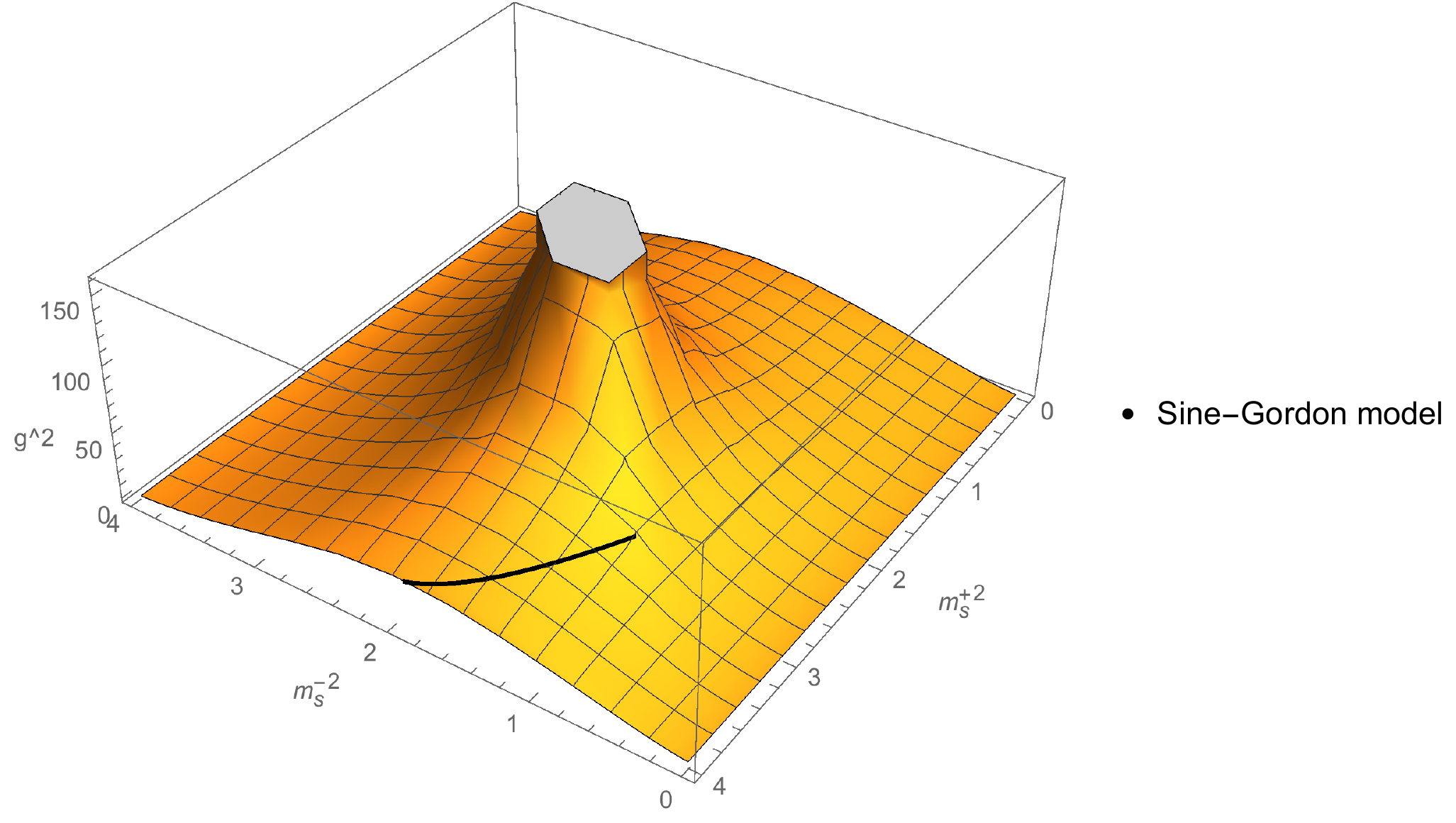}
\caption{\label{fig:3d1}$O(2)$ bound on the coupling to an antisymmetric tensor particle state in the presence of a scalar bound state. Although it is hard to see in the figure, the bound is not smooth along the line $m_s^2+m_a^2=4$, i.e. when the scalar pole and antisymmetric tensor poles collide with each other's cross channel. The solid curve corresponds to the sine-Gordon model with the same spectrum, i.e. for $\pi/2\leq \xi\leq \pi$.}
\end{figure}

In this section we consider numerical bounds in the presence of two bound states. Figure \ref{fig:3d1} corresponds to $O(2)$ and shows the upper bound on the pseudoscalar particle coupling as a function of its mass and that of an extra scalar bound state. The values for the sine-Gordon model lie precisely on the bound surface.

\begin{figure}[htbp]
\centering
\includegraphics[width=0.7\textwidth]{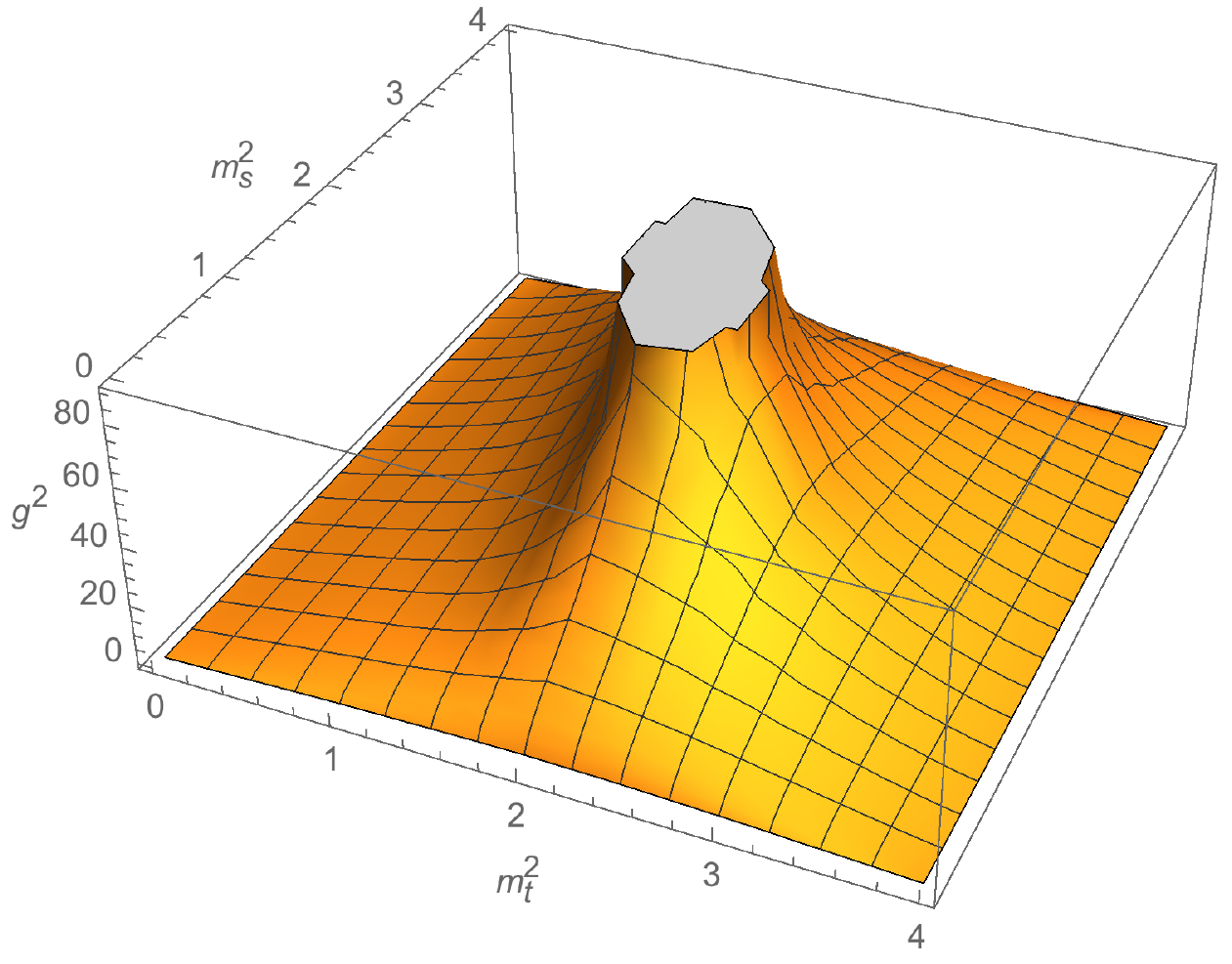}
\caption{\label{fig:SGtensor2}$O(2)$ bound on the coupling to a $(t)$ particle in the presence of a scalar bound state. Exchanging the scalar by a pseudoscalar (i.e. antisymmetric tensor) would yield the same plot.}
\end{figure}

In figure \ref{fig:SGtensor2} we show a bound for the symmetric traceless tensor coupling in the $O(2)$ case in the presence of an extra scalar bound state. Note that for $O(2)$ there is an accidental symmetry between scalar and pseudoscalar states, so repeating the analysis with tensor and pseudoscalar would yield the same results.

\begin{figure}[htbp]
\centering
\includegraphics[width=\textwidth]{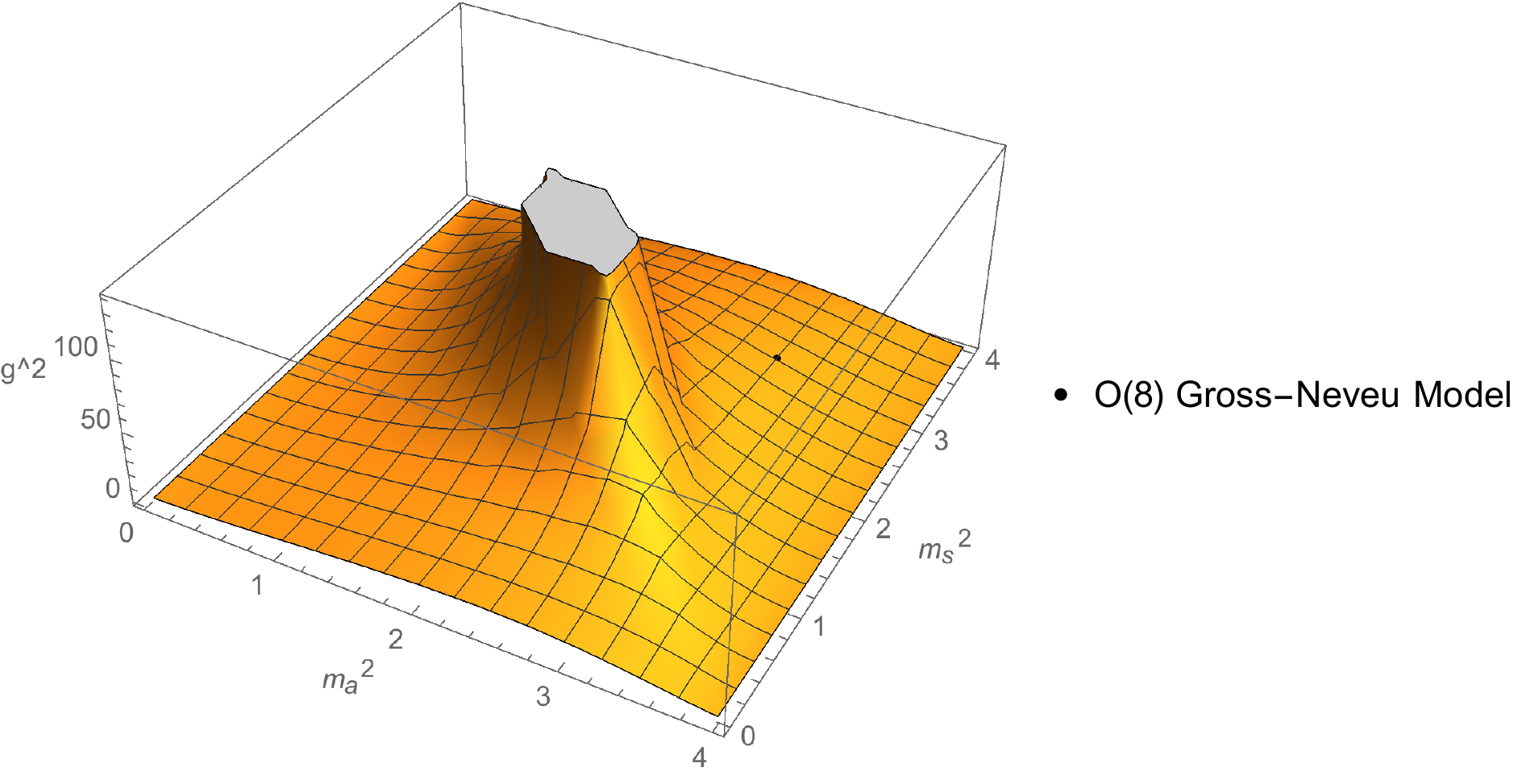}
\caption{\label{fig:GN83d} $O(8)$ bound on  antisymmetric tensor coupling in presence of a scalar bound state. The $O(8)$ Gross-Neveu model is represented by the black point which has degenerate masses.}
\end{figure}

\begin{figure}[htbp]
\centering
\includegraphics[width=.7\textwidth]{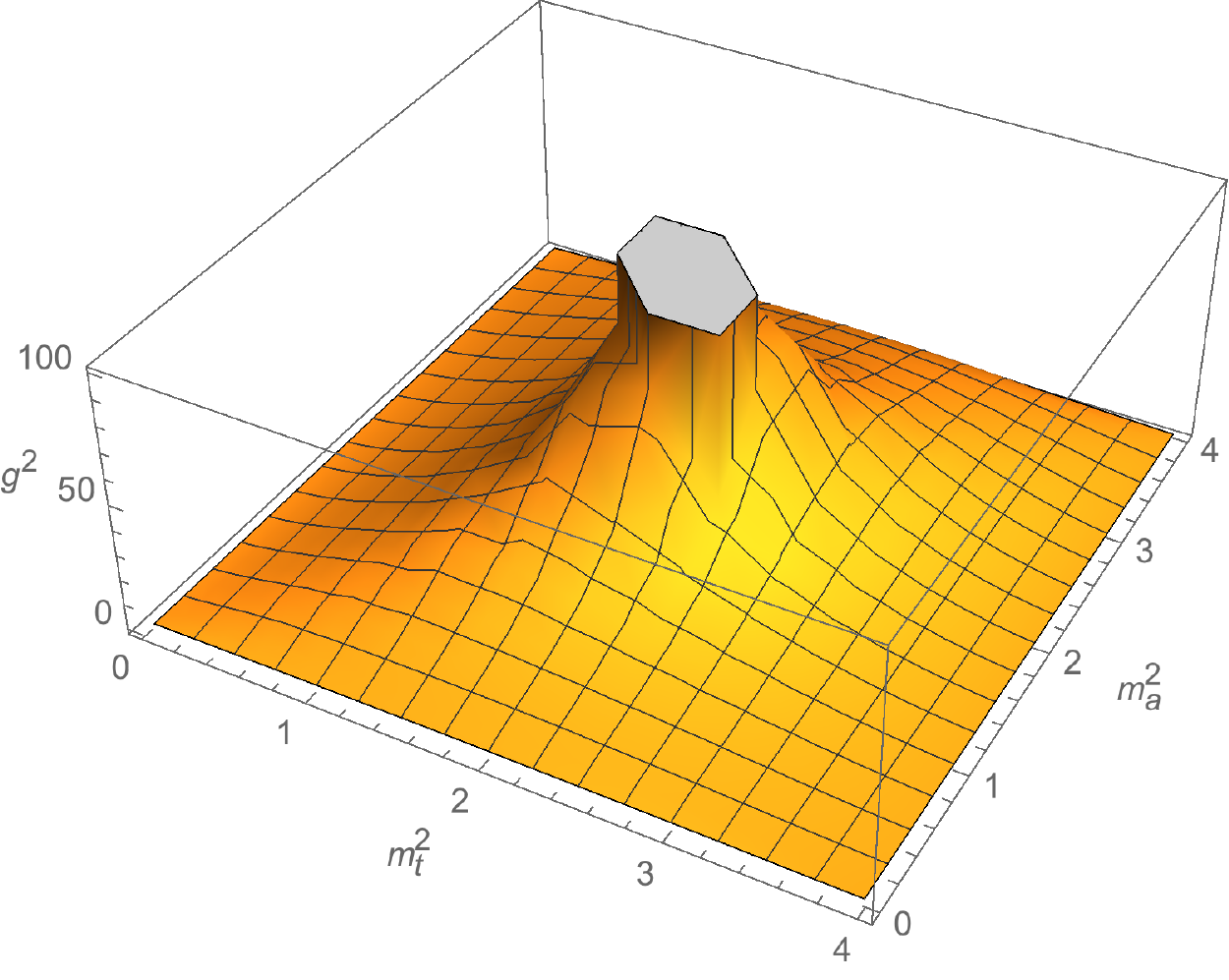}
\caption{\label{fig:GN8tensor}$O(8)$ bound on coupling to a $(t)$ particle in the presence of an extra $(a)$ particle.}
\end{figure}

\begin{figure}[htbp]
\centering
\includegraphics[width=.7\textwidth]{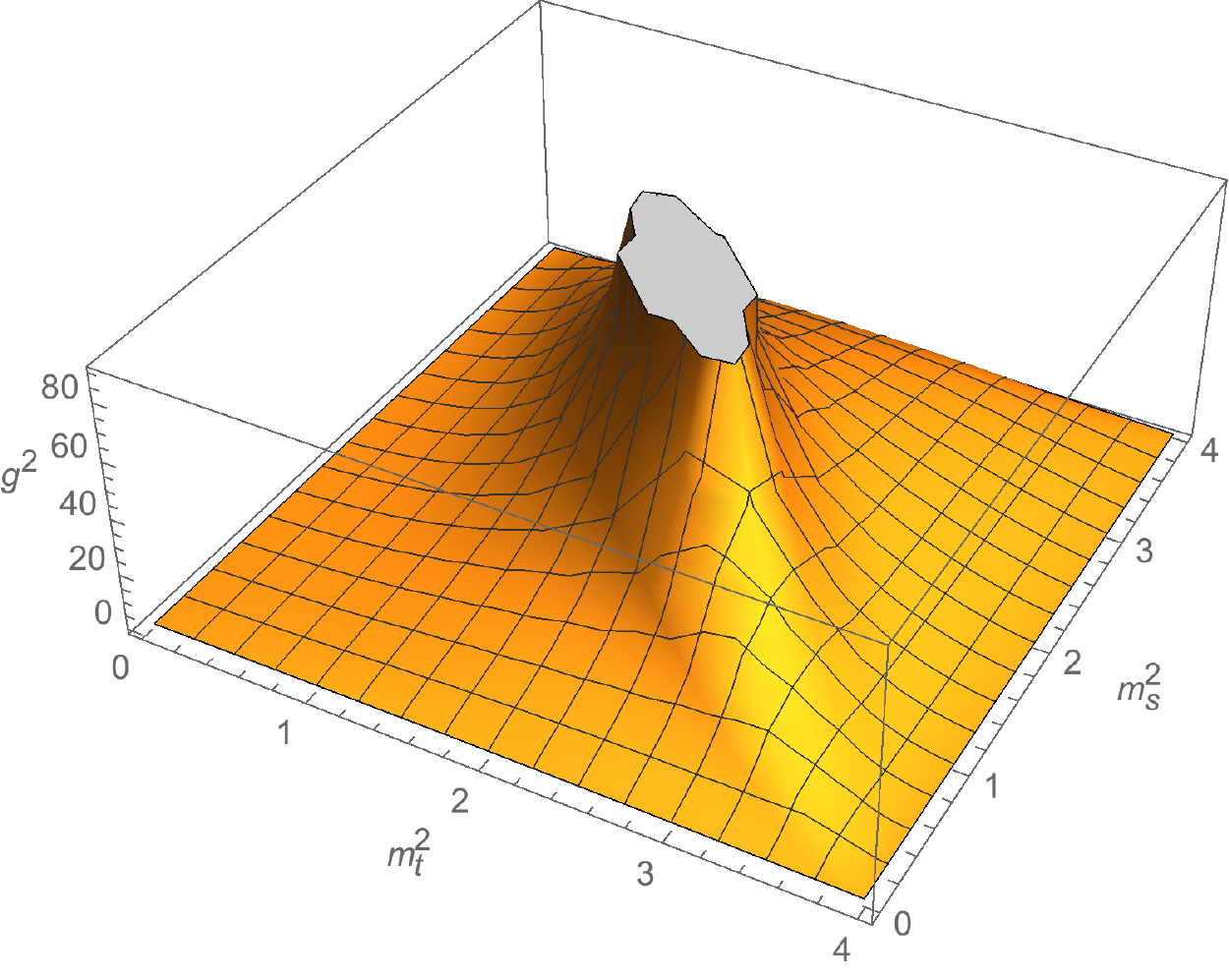}
\caption{\label{fig:GN8scalarp}$O(8)$ bound on coupling to a $(t)$ particle in the presence of an extra $(s)$ particle.}
\end{figure}

In figure \ref{fig:GN83d} we plot an upper bound for the coupling to an antisymmetric tensor $(a)$ particle in $O(8)$ in the presence of an extra bound state in the scalar channel $(s)$. We have marked the $O(8)$ Gross-Neveu model in the plot (it corresponds to having equal masses), which saturates the bound. Figures.~\ref{fig:GN8tensor} and \ref{fig:GN8scalarp} 
repeat the analysis with $(a), (t)$ and $(s),(t)$ bound states respectively.

\clearpage
\bibliographystyle{jhep}
\bibliography{sample}

\end{document}